\documentclass[twoside]{article}
\usepackage{qic}

\usepackage[english]{babel}
\usepackage{amsmath}
\usepackage{amsfonts}
\newcommand{\bra}[1]{\langle #1 |}
\newcommand{\ket}[1]{| #1 \rangle}
\newcommand{\braket}[1]{\langle #1 \rangle}

\renewcommand{\thefootnote}{\fnsymbol{footnote}}  

\begin{document}
\setlength{\textheight}{8.0truein}    

\runninghead{Spatial Search using Flip-flop Quantum Walk}
            {Abhijith J. and A. Patel}

\normalsize\textlineskip
\thispagestyle{empty}
\setcounter{page}{1}

\copyrightheading{0}{0}{2017}{000--000}

\vspace*{0.88truein}

\alphfootnote

\fpage{1}

\centerline{\bf
SPATIAL SEARCH ON GRAPHS WITH MULTIPLE TARGETS}
\vspace*{0.035truein}
\centerline{\bf USING FLIP-FLOP QUANTUM WALK}
\vspace*{0.37truein}
\centerline{\footnotesize
ABHIJITH J.\footnote{E-mail: abhijithj@iisc.ac.in} ~~and
APOORVA PATEL\footnote{E-mail: adpatel@iisc.ac.in}}
\vspace*{0.015truein}
\centerline{\footnotesize\it Centre for High Energy Physics,
Indian Institute of Science}
\baselineskip=10pt
\centerline{\footnotesize\it Bangalore 560012, India}
\vspace*{0.225truein}
\publisher{(received date)}{(revised date)}

\vspace*{0.21truein}

\abstracts{
We analyse the eigenvalue and eigenvector structure of the flip-flop quantum
walk on regular graphs, explicitly demonstrating how it is quadratically
faster than the classical random walk.
Then we use it in a controlled spatial search algorithm with multiple target states, and determine the oracle complexity as a function of the spectral gap and the number of target states.
The oracle complexity is optimal as a function of the graph size and the number of target states, when the spectral gap of the adjacency matrix is $\Theta(1)$.
It is also optimal for spatial search on $D>4$ dimensional hypercubic lattices.
Otherwise it matches the best result available in the literature, with a much simpler algorithm.
Our results also yield bounds on the classical hitting time of random walks on regular graphs, which may be of independent interest.
}{}{}

\vspace*{10pt}

\vspace*{3pt}
\communicate{to be filled by the Editorial}

\vspace*{1pt}\textlineskip    

\setcounter{footnote}{0}
\renewcommand{\thefootnote}{\alph{footnote}}


\section{Introduction}

The quantum search algorithm for an unstructured database, introduced by Lov Grover \cite{grover1996fast}, demonstrated a quadratic speedup over classical search methods.
This seminal work on quantum search has been extended in various directions, and has  emerged as an important primitive in the design of numerous quantum algorithms \cite{grover2015quantum}.
Quantum graph search (QGS) algorithms are extensions of Grover search, with an added locality constraint provided by the spatial structure of the underlying database.
Alternatively, Grover's algorithm is equivalent to quantum search on a complete graph.

Similar to the evolution operator in Grover's algorithm, the QGS operator is a product of two reflection operators.
When the algorithm is interpreted as Hamiltonian evolution, one operator is the oracle that attracts the quantum amplitude towards the target vertices, and the other is a diffusion operator (executed as a quantum walk) that spreads the quantum amplitude around the graph.
Quantum walks are quantum mechanical counterparts of classical random walks, and can be constructed in discrete time \cite{ambainis2001one,aharonov2001quantum} as well as continuous time \cite{farhi1998quantum}.
Both varieties of quantum walks have been used to design search algorithms with provable quantum speedups over their classical versions \cite{childs2004spatial,ambainis2007quantum,magniez2007quantum}. 

In this work, we analyse the QGS algorithm based on the discrete time quantum walk with a coin space considered in Refs.~\cite{shenvi2003quantum,ambainis2005coins}.
First we derive rigorous connections between the spectral decomposition of this quantum walk, dubbed flip-flop quantum walk, and its classical counterpart.
Then we use these results to extend the analysis in Refs.~\cite{ambainis2005coins,tulsi2008faster} to the case of multiple target states, for all regular graphs with a large enough spectral gap.

Similar results have been obtained in the quantum walk framework introduced by Szegedy \cite{szegedy2004quantum}, and an efficient algorithm for quantum search on arbitrary graphs has been proposed in that framework by Krovi et al. \cite{krovi2016walk}.
We compare the results of our QGS algorithm with those of Krovi et al. at the end of Section \ref{sec:QGSanalysis}.
We point out that the Hilbert space of the flip-flop quantum walk has dimension $dN$, which is much smaller than the Hilbert space dimension $N^2$ considered originally by Szegedy.
It is known that for the case of regular graphs the reduction of space complexity from $O(N^2)$ to $O(dN)$ is achievable by constructing an equivalent coined walk \cite{wong2016walk}. 

This article is organized as follows.
Section \ref{sec:prelim} sets up the Quantum Graph Search problem, and contains the notation and definitions used in this work. 
In Section \ref{sec:eigenWU}, we derive two theorems that relate the eigen-properties of the quantum operators considered by us and their classical counterparts.
Theorem \ref{theo_1} relates the eigenvalues and eigenvectors of the quantum flip-flop walk operator to those of the adjacency matrix of the graph.
Theorem \ref{theo_2} relates the eigenvalues and eigenvectors of the quantum search operator to that of the classical random walk on the graph with the target vertices removed.

In Section 4, we derive the main set of equations, Eq.\eqref{eq:eval_eq}, that we use throughout this work to analyse the QGS algorithm.
We also introduce Tulsi's controlled spatial search technique \cite{tulsi2008faster}, and expand our framework to cover the extra states.
Section \ref{sec:QGSanalysis} extends the analysis of Refs.~\cite{ambainis2005coins,tulsi2008faster} to spatial search on regular graphs with multiple targets, using the results from the previous Section.
In particular, we show how the oracle complexity and the success probability of the algorithm depend on the spectral gap of the graph and the number of targets.
In Section \ref{sec:QGScomplete}, we specialize to the exactly solvable case of the QGS algorithm on a complete graph, and compare our results with the simpler analysis of Grover search.
Section \ref{sec:QGSlattice} analyses the special case of the QGS algorithm on a hypercubic lattice in dimension $D>2$, and uses the entire spectral decomposition of the walk operator to derive tighter bounds on the performance of the algorithm.
Our results generalize those obtained in Refs.~\cite{ambainis2005coins,tulsi2008faster} for up to two target states to multiple number of target states and higher dimensions.

Although our focus is on the spatial search problem, several of our results are interesting in their own right.
In Section \ref{sec:Quantum_bounds}, we use our results to obtain bounds on the classical hitting time of random walks on regular graphs by strengthening a result by Szegedy.
Finally, six appendices contain several auxiliary results needed in our analysis.

\section{Preliminaries and Notation}
\label{sec:prelim}

\subsection{Graph properties}

We consider a $d$-regular, undirected graph $G(V,E)$.
Here $V$ is the vertex set and $E$ is the edge set of the graph.
We denote the size of $V$ by $N$, i.e. $|V|=N$, while the size of $E$ is $|E|=dN/2$.
The adjacency matrix $A$ of this graph is an $N \times N$ matrix that encodes the connectivity information of the graph.
We use the normalized adjacency matrix for our work:
\begin{equation}
A_{ij} = \begin{cases}
         \frac{1}{d} ~,~~ \text{if}~ i,j \in E,\\
         0 ~,~~ \text{otherwise}.
         \end{cases}		
\end{equation}
$A$ is a real symmetric matrix. So all its eigenvalues are real, all its eigenvectors can be chosen to be real and they form an complete orthonormal set.
With our normalization, the eigenvalues of $A$ lie in the interval $[-1,1]$ and hence can be expressed as $\cos(\phi)$ for some $\phi \in [0,\pi]$.
We call an angle $\phi$ an eigenphase of $A$.
This adjacency matrix is related to the discrete Laplacian for the graph,
\begin{equation}
\Delta = A - \mathbb{I} ~.
\end{equation}
The largest eigenvalue of $A$ is always $1$, and the corresponding eigenvector is the uniform superposition vector.

We use $E(A,B)$ to denote the set of edges between the sets $A \in V$ and $B \in V$.
Also we label the neighborhood of $u$ as $N(u) = \{v | v\in V, (u,v)\in E \}$. 

\subsection{Flip-flop quantum walk}

The quantum walk is a unitary operator that spreads the quantum amplitude through the graph, while respecting the locality property of the graph.
Physically, locality is demanded by the constraint of relativity, which requires that no signal can travel faster than the speed of light.
Previously, such operators have been labeled $Z$-local in Ref.~\cite{aaronson2003quantum} and ultralocal in Ref.~\cite{patel2005quantum}.

There are many ways to construct local walk operators on $G$; we here follow the flip-flop walk prescription first introduced in Ref.~\cite{shenvi2003quantum}.
The flip-flop walk operator $W$ is defined in the Hilbert space $\mathbb{C}^{d} \otimes \mathbb{C}^{N}$ of dimension $dN$.
The $\mathbb{C}^{d}$ space is often called the coin space attached to every vertex of the graph.
It is spanned by the states $\ket{h}, h \in H$, where $H$ is the set $\{0,1, \ldots , d-1\}$.
The states $\ket{h}\ket{u}$, with $h \in H$ and $u \in V$, form a complete orthonormal vertex state basis for the Hilbert space.

Explicitly,
\begin{equation}
W = SC,
\end{equation}
where $S$ and $C$ are reflection operators, called the shift operator and the coin operator respectively. 
Essentially, $S$ acts only in the graph space and $C$ acts only in the coin space.

\paragraph{Definition of $C$:}
Let $\ket{H} = \frac{1}{\sqrt{d}} \sum_{h \in H} \ket{h}$ be the uniform superposition of the coin states.
Then the coin operator is $C = 2P_H - \mathbb{I}$, where $P_H = \ket{H}\bra{H} \otimes \mathbb{I}$.
It is also refered to as the Grover coin, and $C^2=\mathbb{I}$.
It is possible to construct other unitary coins that mix the coin degrees of freedom, but we shall not consider them.

\paragraph{Definition of $S$:}
The edges emanating from each vertex are associated with the basis vectors of the coin space.
In some graphs (e.g. Cayley graphs), there exists a natural way to map the edges to its corresponding basis state.
For a general graph, however, such a mapping may not exist.
So we define a function $f: E \rightarrow H \times H$, which maps every edge at a vertex to its corresponding basis state in the coin space (such a function can be easily constructed by parsing $E$).
For an edge $e = (u,v)$, $f((u,v)) = (h,g)$ means that the edge $e$ is mapped to the basis state $\ket{h}$ in the coin space of $u$ and to the basis state $\ket{g}$ in the coin space of $v$.
From the definition of $f$ it is obvious that if $f((u,v))=(h,g)$ then $f((v,u))=(g,h)$.

The dimension $dN$ of the Hilbert space is twice the number of edges in the graph.
With every edge $e=(u,v) \in E$, we associate two states:
\begin{equation}
\label{eplus}
\ket{e^+} = \frac{1}{\sqrt{2}}(\ket{h}\ket{u} + \ket{g}\ket{v}), 
\end{equation} 
\begin{equation} 
\label{eminus}
\ket{e^-} = \frac{1}{\sqrt{2}}(\ket{h}\ket{u} - \ket{g}\ket{v}),
\end{equation}
where $f((u,v)) = (h,g)$.
These states form a complete orthonormal edge state basis for the Hilbert space.
The shift operator $S$ is a diagonal matrix in this edge basis:
\begin{equation}
\label{eq:def_S}
S = \sum_{e \in E} \ket{e^+} \bra{e^+} - \ket{e^-} \bra{e^-} ~.
\end{equation}
$S$ acts on the Hilbert space states as $S \ket{h,u} = \ket{g,v}$.
We can write $S$ also as:
\begin{equation}
S = \sum_{\{(u,v) \in E : f((u,v)) = (h,g)\}} \ket{h}\bra{g} \otimes \ket{u} \bra{v}.
\end{equation}
Clearly, $S$ is a reflection operator, with $S^2 = \mathbb{I}$. 

The $C$ and $S$ operators are not arbitrarily defined.
They are the quantum extensions of the two elementary steps that a classical walker on $G$ performs.
First the classical walker rolls a $d$-dimensional coin while positioned at one vertex.
The quantum version of this is the application of $C$.
Next the classical walker moves to a neighboring vertex based on the outcome of the roll of the coin.
This is captured in $S$, which is responsible for entangling the coin and the vertex degrees of freedom.
It is then natural to look for a more quantifiable connection between a classical random walk and a quantum walk.
We derive such a connection in Theorem 1.   

\subsection{The search problem}

In spatial search on $G$, we are given a set $T \subset V$, containing $M$ target vertices.
We have to find the location of any of the $M$ vertices, given oracular access to the elements of $T$.
The oracle is defined in the usual way, as the reflection operator:
\begin{equation}
O = \mathbb{I} - 2 \sum_{i \in T} \ket{\psi_i}\bra{\psi_i} = \mathbb{I} - 2P ~.
\end{equation}
Here $\ket{\psi_u} = \frac{1}{\sqrt{d}} \sum_{h \in H} \ket{h,u}$ is the uniform superposition state over all the coin states at the vertex $u$.
The spatial search operator is defined as $U = WO$.
Since $C$, $S$ and $O$ are all real reflection operators, both $W$ and $U$ are orthogonal operators.
Consequently, eigenvalues of $W$ and $U$ are either real or come in complex conjugate pairs.

In what follows, we demonstrate that we can reach the vertices in $T$ with high probability, by repeatedly applying $U$ to a starting state independent of $T$.  
\section{Eigenvalues and Eigenvectors of $W$ and $U$}
\label{sec:eigenWU}
 
Now we state and prove two important theorems that connect the properties of the classical random walk on $G(V,E)$ defined by $A$, to the flip-flop quantum walk $W$ on the same graph.
Let $\ket{\Phi_k}$ to be an eigenvector of $W$ with eigenvalue $e^{i \phi_k}$.
For every vertex, $i \in V$, we define $a_{ki} = \braket{\Phi_k|\psi_i}$.
These coefficients provide the connection between the eigenvectors of $W$ and $A$ as follows.%
\footnote{We use the vector sign to indicate eigenvectors of the classical walk operator (e.g. $\vec{a}$ for $A$) and the Dirac notation to indicate eigenvectors of the quantum walk operator (e.g. $\ket{\Phi}$ for $W$).}

\smallskip
\begin{theorem}
\label{theo_1}
The vector $\vec{a}_k = (a_{k1}, a_{k2},....)$ is an eigenvector of the adjacency matrix $A$ of $G$ with eigenvalue $\cos\phi_k$ when $0<\phi_k<\pi$  (i.e. for all the eigenvectors of $W$ which do not have eigenvalues $\pm1$). 

(\textbf{Converse}) Also, for every eigenvector of $A$ with eigenvalue $\cos(\phi)$ and $0 < \phi_k < \pi$, there exists two eigenvectors of $W$ with eigenvalues $e^{i\phi}$ and $e^{-i\phi}$. 
\end{theorem}

\noindent{\bf Proof:}
The $\ket{e^+}$ vectors defined in Eq.\eqref{eplus} for every edge of the graph are eigenvectors of $S$ with eigenvalue $1$. Therefore,
\begin{eqnarray}
\braket{e^+ |W| \Phi_k} = e^{i \phi_k} \braket{e^+ | \Phi_k}
  &=& \braket{e^{+} | SC | \Phi_k} \\
  &=& 2 \sum_{i \in V} \braket{e^+|\psi_i}\braket{\psi_i|\Phi_k} -\braket{e^+ |\Phi_k} ~.
\end{eqnarray}
Let the nodes $u$ and $v$ be the ends of the edge $e$ in this equation.
Then $\ket{e^+}$ has non-zero overlap only with $\ket{\psi_u}$ and $\ket{\psi_v}$.
So we can solve for $\braket{e^+ |\Phi_k}$:
\begin{equation}
\braket{\Phi_k|e^+} = \sqrt{\frac{2}{d}} ~\frac{a_{ku}+a_{kv}}{1 + e^{-i\phi_k}} ~.
\end{equation} 

Similarly, the $\ket{e^-}$ vectors defined in Eq.\eqref{eminus} are eigenvectors of $S$ with eigenvalue $-1$, and we obtain:
\begin{equation}
\braket{\Phi_k|e^-} = \sqrt{\frac{2}{d}}~ \frac{a_{ku}-a_{kv}}{1 - e^{-i\phi_k}} ~.
\end{equation} 
There is a sign ambiguity in this expression that arises from the definition of $\ket{e^-}$, but it will not reflect in the final results.

Note that since $\ket{e^+},\ket{e^-}$ form a complete basis, all the components of $\ket{\Phi_k}$ are determined in terms of $a_{ku}$ and the $\phi_k$.
We have
\begin{align}
a_{ki} &= \braket{\Phi_k| \psi_i} \\
& = \sum_{e = (u,v) \in E}  \braket{\Phi_k| e^+}\braket{e^+| \psi_i}+ \braket{\Phi_k| e^-}\braket{e^-| \psi_i} \\
& =  \frac{1}{d} \sum_{j \in N(i)} \left( \frac{a_{ki} + a_{kj}}{1 + e^{-i\phi_k}} + \frac{a_{ki} - a_{kj}}{1 - e^{-i \phi_k}} \right) ~.
\end{align}
Using the fact that $\ket{\psi_i}$ are eigenvectors of $C$ with eigenvalue $1$, we can also express $a_{ki}$ as
\begin{align}
a_{ki} &= \braket{\Phi_k| C | \psi_i} \\
       &= e^{-i\phi_k} \braket{\Phi_k| S | \psi_i} \\
       &= e^{-i\phi_k} \sum_{e=(u,v) \in E} \left( \braket{\Phi_k| e^+}\braket{e^+| \psi_i} - \braket{\Phi_k| e^-}\braket{e^-| \psi_i} \right) \\
       &= \frac{e^{-i\phi_k}}{d} \sum_{j \in N(i)} \left( \frac{a_{ki} + a_{kj}}{1 + e^{-i \phi_k}} - \frac{a_{ki} - a_{kj}}{1 - e^{-i \phi_k}} \right) ~.
\end{align}

Equating these two expressions for $a_{ki}$, we get
\begin{equation}
\tan(\frac{\phi_k}{2}) \sum _{j \in N(i)} (a_{ki} + a_{kj}) = \cot(\frac{\phi_k}{2}) \sum _{j \in N(i)} (a_{ki} - a_{kj}) ~.
\end{equation}
Since $(\cot{\phi_k/2}-\tan{\phi_k/2}) = 2\cot{\phi}$, and
$(\tan{\phi_k/2}+\cot{\phi_k/2}) = 2/\sin{\phi}$, we can simplify
\begin{equation}
\frac{1}{d} \sum_{j\in N(i)} a_{kj} = a_{ki}~\cos\phi_k ~.
\end{equation}
This can be rewritten as the eigenvalue equation $A\vec{a}_k = \cos\phi_k\vec{a}_k$, proving the first part of the theorem.

\noindent{\bf Proof of converse:}
Let $\vec{a}$ be such that $A\vec{a} = \cos(\phi)\vec{a}$, and let $a_u$ be the component of $\vec{a}$ corresponding to the vertex $u$.
Now consider $\ket{\Phi} \in \mathbb{C}^{Nd}$ defined in the edge basis as follows:
\begin{equation}
\braket{\Phi| e^\pm} = \sqrt{\frac2d} ~\frac{a_u \pm a_v}{1 \pm e^{-i \phi}} ~,
\end{equation}
where $e=(u,v)$.
We also have, for any $u \in V$,
\begin{equation}
\ket{\psi_u} = \frac{1}{\sqrt{d}} \sum_{h \in H} \ket{h,u}
 = \frac{1}{\sqrt{2d}} \sum_{w \in N(u), e = (u,w)} (\ket{e^+} + \ket{e^-}) ~.
\end{equation}
Hence, it follows that
\begin{align}
\braket{\psi_u|\Phi} &= 
\frac{1}{\sqrt{2d}}\sum_{w \in N(u), e = (u,w)} (\braket{e^+|\phi} + \braket{e^-|\phi}) ~, \\
 & = \frac{1}{d} \sum_{w \in N(u)} \left( \frac{a_u + a_w}{1 + e^{i\phi}} +  \frac{a_u - a_w}{1 - e^{i \phi}} \right) ~, \\
 &= a_u \left( \frac{1}{1 + e^{i\phi}} + \frac{1}{1 - e^{i\phi}} \right) + \cos(\phi)~a_u \left( \frac{1}{1 + e^{i\phi}} - \frac{1}{1 - e^{i\phi}} \right) ~.
\end{align}
In the last step, we have used the eigenvalue equation for $A$, i.e. $\frac{1}{d} \sum_{w \in N(u)} a_w = \cos(\phi)~a_u$.
Simplifying the last line, we get the expected result, 
\begin{equation}
\braket{\psi_u| \Phi} = a_u ~.
\end{equation} 
 
Next we compute the matrix elements $\braket{e^\pm|W|\Phi}$, using the fact that  $S\ket{e^\pm}=\pm\ket{e^\pm}$:
\begin{align}
\label{eWPhi_first}
\braket{e^\pm|W|\Phi} = \braket{e^\pm|SC|\Phi}
  &= \pm\braket{e^\pm|C|\Phi}, \\
  &= \pm\braket{e^\pm| 2\sum_{i \in V} \ket{\psi_i} \bra{\psi_i} - \mathbb{I} |\Phi} ~, \\
  &= \pm\sqrt{\frac{2}{d}} (\braket{\psi_u|\Phi} \pm \braket{\psi_v|\Phi}) \mp \braket{e^\pm|\Phi} ~, \\
  &= \pm\sqrt{\frac{2}{d}} (a_u \pm a_v) \mp \sqrt{\frac{2}{d}}~\frac{a_u \pm a_v}{1 \pm e^{i\phi}} ~, \\
&= e^{i\phi} \sqrt{\frac{2}{d}} ~\frac{a_u \pm a_v}{1 \pm e^{i\phi}} ~, \\
&= e^{i\phi} \braket{e^\pm|\Phi} ~.
\label{eWPhi_last}
\end{align}
Together, they imply $W\ket{\Phi} = e^{i\phi}\ket{\Phi}$.
Taking complex conjugates, we get for $0>-\phi>-\pi$, $W\ket{\Phi^*} = e^{-i\phi}\ket{\Phi^*}$.
\hfill $\square\,$ 

\smallskip
From this theorem, it is clear that the $a_{ku}$ being components of an eigenvector of $A$ can always chosen to be real, and we do so.
Once the eigenphase $\phi$ is chosen, the $a_{ku}$ values completely determine the corresponding eigenvectors of $W$.
Note that we have to explicitly exclude the values $\phi=0,\pi$, because the denominators $1\pm e^{i\phi}$ appear in the proof.

\smallskip
\begin{corollary}
\label{cor:1}
If $G$ is non-bipartite, $W$ has $2N-2$ complex eigenvalues.
If $G$ is bipartite, $W$ has $2N-4$ complex eigenvalues. 
\end{corollary}

The spectra of adjacency matrices associated with graphs are well studied in the context of spectral graph theory.
It is well-known that the normalized adjacency matrix of a graph has the unique largest eigenvalue $1$, when it is made up of a single connected component.
Also the normalized adjacency matrix has a $-1$ eigenvalue, if and only if the graph is bipartite \cite{chung1996spectral}.
So, when $G$ is a non-bipartite graph, the adjacency matrix has $N-1$ eigenvalues in the interval $(1,-1)$.
Then by Theorem \ref{theo_1}, $W$ has only $2(N-1)$ complex eigenvalues.
The rest of the eigenvalues are either $1$ or $-1$ (their exact numbers are counted in Appendix A).
Similarly, for bipartite graphs the adjacency matrix has $N-2$ eigenvalues in the interval $(-1,1)$, which implies that $W$ has $2N-4$ complex eigenvalues.
\hfill $\square\,$ 

\smallskip
We denote by $\ket{\Phi_0}$ the uniform superposition state $\frac{1}{\sqrt{N}}\sum_{v \in V}\ket{H,v}$, which is also the starting state of our search algorithm.
This state is an eigenvector of $W$ with eigenvalue $1$.
For non-bipartite graphs, we show in Appendix C that the search algorithm preserves the subspace spanned by $\ket{\Phi_0}$ and the eigenvectors corresponding to the complex eigenvalues.

For bipartite graphs with the bipartition $(F,\overline{F})$ of $V$, the state
\begin{equation}
\ket{\Phi_b} = \frac{1}{\sqrt{N}} (\sum_{v \in F} \ket{H,v} - \sum_{v \in \overline{F}} \ket{H,v}),
\end{equation}
is an eigenvector of $W$ with eigenvalue $-1$.
We show in Appendix C that the invariant subspace includes this vector also.
All other eigenvectors of $W$ with real eigenvalues turn out to be irrelevant for the search algorithm.  

\paragraph{\bf Definition of leaking walk matrix \cite{szegedy2004quantum}:}
Given the set $T \subset V$ of the target states, the leaking walk matrix $\tilde{A}_T$ is the matrix obtained by removing all rows and columns corresponding to $T$ from $A$.
$\tilde{A}$ is a real symmetric matrix; its eigenvalues are therefore real and we choose its eigenvectors to be real as well.
$\tilde{A}_T$ is not a stochastic matrix, however, so its eigenvalues are strictly less than $1$ in magnitude.
The eigenvalues of the leaking walk matrix govern the evolution of the error probability, i.e. the probability that the walker is still at one of the non-target vertices, of a classical random walk with perfect sinks at the target vertices.

Since $O$ acts trivially on $V-T$, $U$ restricted to $V-T$ has the same effect as $W$.
So, just like the relation between eigenvalues of $A$ and $W$, we anticipate a relation between eigenvalues of $\tilde{A}_T$ and $U$.
We now prove such a relation, being careful to take in to account the amplitude exchanges between $V-T$ and $T$.

\begin{theorem}
\label{theo_2}
For every eigenvector $\ket{\Lambda}$ of $U$ with eigenvalue $e^{i\lambda}$ and $0 < \lambda < \pi$, the projected vector $\vec{\Lambda} = (\braket{\Lambda|\psi_i} | i \in V-T)$ is an eigenvector of $\tilde{A}_T$ with eigenvalue $\cos(\lambda)$.

(\textbf{Converse}) Also, for every eigenvector of $\tilde{A}_T$ with eigenvalue $\cos(\lambda)$, there exist two eigenvectors of $U$ with eigenvalues $e^{i\lambda}$ and $e^{-i\lambda}$.
\end{theorem}

\noindent{\bf Proof:}
Note that $U = SCO$ is real; so $U\ket{\Lambda} = e^{i\lambda}\ket{\Lambda}$ implies $U\ket{\Lambda^*} = e^{-i\lambda}\ket{\Lambda^*}$.
Also,
\begin{equation}
\label{COident}
CO = (2\sum_{i \in V} \ket{\psi_i}\bra{\psi_i} - \mathbb{I})
     (\mathbb{I} - 2\sum_{i \in T} \ket{\psi_i}\bra{\psi_i})
   = 2\sum_{i \in V-T} \ket{\psi_i}\bra{\psi_i} - \mathbb{I} ~.
\end{equation}
Therefore,
\begin{align}
\braket{e^+|U|\Lambda} = e^{i\lambda}\braket{e^+|\Lambda}
  &= \braket{e^{+} | SCO | \Lambda} \\
  &= \braket{e^{+} | 2\sum_{i \in V-T} \ket{\psi_i}\bra{\psi_i}-\mathbb{I} | \Lambda} \\
  &= 2\sum_{i \in V-T} \braket{e^+|\psi_i}\braket{\psi_i|\Lambda} - \braket{e^+ |\Lambda} ~.
\end{align}

There are three types of edges to consider:
\begin{enumerate}
\item Edges that are completely in $V-T$. For these we get, 
\begin{equation}\label{eq:edge_overlap}
\braket{\Lambda| e^+} = \sqrt{\frac{2}{d}} ~\frac{\vec{\Lambda}_u + \vec{\Lambda}_v}{1 + e^{-i\lambda}} ~.
\end{equation}
\item Edges that go between $V-T$ and $T$. Then we find,
\begin{equation}
\braket{\Lambda|e^+} = \sqrt{\frac{2}{d}} ~\frac{\vec{\Lambda}_u}{1 + e^{-i\lambda}} ~,
\end{equation}
where $u$ is the end of the edge in $V-T$.
\item Edges that are completely in $T$. Here we have,
\begin{equation}
\label{no_overlap+}
\braket{\Lambda|e^+} = 0 ~.
\end{equation}
\end{enumerate}

Similarly, evaluating $\braket{e^-|U|\Lambda}$, we obtain:
\begin{enumerate}
\item For edges that are completely in $V-T$,
\begin{equation}
\braket{\Lambda|e^-} = \sqrt{\frac{2}{d}} ~\frac{\vec{\Lambda}_u - \vec{\Lambda}_v}{1 - e^{-i \lambda}} ~.
\end{equation}
\item For edges that go between $V-T$ and $T$,
\begin{equation}
\braket{\Lambda|e^-} = \sqrt{\frac{2}{d}} ~\frac{\vec{\Lambda}_u}{1 - e^{-i \lambda}} ~.
\end{equation}
where $u$ is the end of the edge in $V-T$.
\item For edges that are completely in $T$,
\begin{equation}
\label{no_overlap-}
\braket{\Lambda|e^-} = 0 ~.
\end{equation}
\end{enumerate}

As in case of Theorem \ref{theo_1}, we find two independent expressions for $\vec{\Lambda}_i$:
\begin{align}
\vec{\Lambda}_i &= \braket{\Lambda|\psi_i}
  = \sum_{e \in E} (\braket{\Lambda|e^+}\braket{e^+|\psi_i} + \braket{\Lambda|e^-}\braket{e^-|\psi_i}) ~, \\
  &= \frac{1}{d} \sum_{j \in N(i) \cap (V-T) } \left( \frac{\vec{\Lambda}_i + \vec{\Lambda}_j}{1 + e^{-i\lambda}} + \frac{\vec{\Lambda}_i - \vec{\Lambda}_j}{1 - e^{-i\lambda}} \right) + \frac{1}{d} \sum_{j \in N(i) \cap T} \left( \frac{\vec{\Lambda}_i }{1 + e^{-i\lambda}} + \frac{\vec{\Lambda}_i}{1 - e^{-i\lambda}} \right) ~.
\end{align}
 
\begin{align}
\vec{\Lambda}_i &= \braket{\Lambda |OC| \psi_i}
  = e^{-i\lambda} \braket{\Lambda |S| \psi_i} ~, \\
  &= e^{-i\lambda} \sum_{e \in E} (\braket{\Lambda|e^+}\braket{e^+|\Lambda} - \braket{\Lambda|e^-}\braket{e^-|\Lambda}) ~, \\
  &= \frac{e^{-i \lambda}}{d} \sum_{j \in N(i) \cap (V-T) } \left( \frac{\vec{\Lambda}_i + \vec{\Lambda}_j}{1 + e^{-i\lambda}} - \frac{\vec{\Lambda}_i - \vec{\Lambda}_j}{1 - e^{-i\lambda}} \right)  + \frac{e^{-i\lambda}}{d} \sum_{j \in N(i) \cap T} \left( \frac{\vec{\Lambda}_i}{1 + e^{-i\lambda}} - \frac{\vec{\Lambda}_i}{1 - e^{-i \lambda}} \right) ~.
\end{align}

Equating these expressions for $\vec{\Lambda}_i$, we get
\begin{align}
\vec{\Lambda}_i \left( \frac{1 - e^{-i\lambda}}{1 + e^{-i\lambda}} + \frac{1 + e^{-i\lambda}}{1 - e^{-i\lambda}} \right) &= \frac{1}{d} \sum_{j\in N(i) \cap V-T} \vec{\Lambda}_j \left( \frac{1 + e^{-i\lambda}}{1 - e^{-i\lambda}} - \frac{1 - e^{-i\lambda}}{1 + e^{-i\lambda}} \right) ~, \\ 
\vec{\Lambda}_i ( \cot\frac{\lambda}{2} - \tan\frac{\lambda}{2}) &= \frac{1}{d} \sum_{j\in N(i) \cap V-T} \vec{\Lambda}_j ( \tan\frac{\lambda}{2} + \cot\frac{\lambda}{2}) ~.
\end{align}
This result simplifies to:
\begin{equation}
\vec{\Lambda}_i \cos(\lambda) = \frac{1}{d} \sum_{j\in N(i) \cap V-T} \vec{\Lambda}_j ~, 
\end{equation}
which is equivalent to $\tilde{A}_T \vec{\Lambda} = \cos(\lambda)~\vec{\Lambda}$, proving the first part of the theorem.

\noindent{\bf Proof of converse:}
The proof of the converse is again on the same lines as in Theorem \ref{theo_1}.
Let $\vec{\Lambda}$ be such that $\tilde{A}_T \vec{\Lambda} = \cos(\lambda) \vec{\Lambda}$, and let $\Lambda_u$ be the component of $\vec{\Lambda}$ corresponding to the vertex $u$.
Then we define $\ket{\Lambda} \in \mathcal{C}^{Nd}$, in terms of its components in the edge basis, as follows:
\begin{enumerate}
\item For edges completely in $V-T$: 
\begin{equation}
\braket{\Lambda|e^\pm} = \sqrt{\frac{2}{d}} ~\frac{\vec{\Lambda}_u \pm \vec{\Lambda}_v}{1 \pm e^{-i\lambda}} ~.
\end{equation}
\item For edges going between $V-T$ and $T$:
\begin{equation}
\braket{\Lambda|e^\pm} = \sqrt{\frac{2}{d}} ~\frac{\vec{\Lambda}_u }{1 \pm e^{-i\lambda}} ~.
\end{equation}
where $u$ is the end of the edge in $V-T$.

\item For edges completely in $T$,
\begin{equation}
\braket{\Lambda| e^\pm} = 0 ~.
\end{equation}
\end{enumerate}

Now that we have defined $\ket{\Lambda}$, we show that $U\ket{\Lambda} = e^{i\lambda}\ket{\Lambda}$, taking the same steps that were used in the proof of Theorem \ref{theo_1}.
For edges completely in $V-T$, $U$ reduces to $W$ and $\braket{e^\pm|U|\Lambda} = e^{\pm i\lambda}\braket{e^\pm|\Lambda}$, in perfect analogy of obtaining Eq.\eqref{eWPhi_last} from Eq.\eqref{eWPhi_first}.
For edges going between $V-T$ and $T$, with $u$ being the end of the edge in $V-T$, use of Eq.\eqref{COident} gives:
\begin{align}
\braket{e^\pm|U|\Lambda} = \braket{e^\pm|SCO|\Lambda}
  &= \pm\sqrt{\frac{2}{d}} \braket{\psi_u|\Lambda} \mp \braket{e^\pm|\Lambda} ~, \\
  &= \pm\sqrt{\frac{2}{d}} ~\vec{\Lambda}_u \mp \sqrt{\frac{2}{d}}~\frac{\vec{\Lambda}_u}{1 \pm e^{i\lambda}} ~, \\
&= e^{i\lambda} \sqrt{\frac{2}{d}} ~\frac{\vec{\Lambda_u}}{1 \pm e^{i\lambda}} ~, \\
&= e^{i\lambda} \braket{e^\pm|\Lambda} ~.
\end{align}
For edges completely in $T$, $\braket{e^\pm|U|\Lambda} = \mp\braket{e^\pm|\Lambda} = 0$.
All cases put together prove the desired result $U\ket{\Lambda} = e^{i\lambda}\ket{\Lambda}$.
\hfill $\square\,$ 

\smallskip
Next we prove a useful corollary of Theorem \ref{theo_2}, which helps us analyse the search algorithm with multiple target states.
Let $\cos\alpha$ be the largest eigenvalue of $\tilde{A}_T$.
From Theorem \ref{theo_2}, we know that $e^{i\alpha}$ is an eigenvalue of $U$, and let the corresponding eigenvector be $\ket{\alpha}$.

\smallskip
\begin{corollary}
\label{cor:2}
For $i \in T$, define $x_i = \braket{\psi_i|\alpha}$.
Then it is possible to choose these $x_i$ such that all of them are negative imaginary numbers, as long as the removal of $T$ from $V$ does not make the graph disconnected.
\end{corollary}

\noindent{\bf Proof:} 
The $x_i$ are defined only for $i \in T$, while Theorem \ref{theo_2} shows that similar components $\vec{\Lambda}_u$ defined for $V-T$ are equal to the components of the eigenvector of $\tilde{A}_T$.
Now for $i \in T$, 
\begin{equation}
\ket{\psi_i} = \frac{1}{\sqrt{2d}} \sum_{j \in N(i),~e=(i,j)} (\ket{e^+} + \ket{e^-}) ~. 
\end{equation}
So
\begin{equation}
x_i = \braket{\psi_i | \alpha}
    = \frac{1}{\sqrt{2d}} \sum_{j \in N(i),~e=(i,j)} (\braket{e^+ | \alpha} + \braket{e^- | \alpha}) ~.
\end{equation}

In the proof of Theorem \ref{theo_2}, we saw that $\ket{\alpha}$ has zero overlap with the edge states corresponding to edges completely in $T$ (cf. Eqs.\eqref{no_overlap+},\eqref{no_overlap-}).
So 
\begin{align}
x_i &= \frac{1}{\sqrt{2d}} \sum_{j \in N(i) \cap (V-T),~e=(i,j)} (\braket{e^+ | \alpha} + \braket{e^- | \alpha}) ~, \\
    &= \frac{1}{d} \sum_{j \in N(i) \cap (V-T)} \braket{\psi_j|\alpha} ~(\frac{1}{1+e^{i\alpha}} - \frac{1}{1-e^{i\alpha}}) ~, \\
	&=  -\frac{i}{d\sin\alpha} \sum_{j \in N(i) \cap (V-T)} \vec{\alpha}_j ~.
\label{eq:x_express}
\end{align} 
Here $\vec{\alpha}$ is the eigenvector of $\tilde{A}_T$ with eigenvalue $\cos\alpha$, whose components are equal to $\braket{\psi_j|\alpha}$ for $j \in V-T$ according to Theorem \ref{theo_2}.
Also, the minus sign in the second line arises from the reversal of the edge orientation.

When removal of the target states does not make the graph disconnected, $\tilde{A}_T$ is irreducible.
For an irreducible and non-negative matrix, the Perron-Frobenius theorem guarantees that the eigenvector corresponding to the largest eigenvalue can be chosen to have all real and strictly positive components \cite{meyer2000matrix}.
With such a choice, all the $x_i$'s are negative imaginary numbers.
\hfill $\square\,$ 

In what follows, we use two different summation conventions for the eigenvalue label $k$.
The $\sum_{k\ne0}$ is used where eigenvectors explicitly appear.
It denotes the sum over the all complex eigenvectors $\ket{\Phi_k}$, their complex conjugates $\ket{\Phi_k^*}$, and $\ket{\Phi_b}$ if it exists.
The $\sum_{k>0}$ denotes the sum over all eigenvalues of $A$ not equal to $1$.
When $A$ comes from a bipartite graph, then it includes the terms that come with the $-1$ eigenvalue with weight $\frac{1}{2}$ (to account for the fact that $\ket{\Phi_b}$ does not have a complex conjugate). 

\section{The smallest non-zero eigenphase of U}

Theorem \ref{theo_2} gives an expression for the eigenvalues and the eigenvectors of $U$ in terms of those of $\tilde{A}_T$.
But $\tilde{A}_T$ is often as hard to work with as $U$.
So we now derive an alternate expression for the eigenvalues of $U$.
To analyse our search algorithm, we will only need the smallest non-zero eigenphase of $U$, so we specialize this new expression to that case.

We start by deriving an equation for the components of $\ket{\Lambda}$ in the eigenbasis of $W$.
\begin{align}
\braket{\Phi_k|U|\Lambda} =  e^{i \lambda} \braket{\Phi_k|\Lambda}
  &= \braket{\Phi_k|WO|\Lambda} ~, \\
  &= e^{i \phi_k} \bra{\Phi_k} \mathbb{I} - 2 \sum_{i \in T} \ket{\psi_i}\bra{\psi_i} \ket{\Lambda} ~,
\end{align}
gives
\begin{equation}
\braket{\Phi_k|\Lambda} = \frac{2}{1-e^{i(\lambda-\phi_k)}} \sum_{i \in T} \braket{\Phi_k|\psi_i}\braket{\psi_i|\Lambda} ~.
\end{equation}
Now, in terms of the complete orthonormal basis $\{|\Phi_k\rangle\}$ and $j \in T$, consider
\begin{align}
\braket{\psi_j|\Lambda} &= \sum_{k}\braket{\psi_j|\Phi_k}\braket{\Phi_k|\Lambda},\\
&= \sum_{i \in T}\sum_k \braket{\psi_j|\Phi_k}\frac{2}{1-e^{i(\lambda-\phi_k)}} \braket{\Phi_k|\psi_i}\braket{\psi_i|\Lambda} ~.
\end{align}
Using the identity $\frac{2}{1-e^{i\theta}} = 1+i\cot\frac{\theta}{2}$, this can be converted to
\begin{equation}
\sum_{i \in T}\sum_k \braket{\psi_j|\Phi_k}\cot(\frac{\lambda-\phi_k}{2})\braket{\Phi_k|\psi_i}\braket{\psi_i|\Lambda} = 0 ~.
\end{equation}
This expression holds for any eigenphase of $U$.

Specializing to the smallest eigenphase $\alpha$, we have $x_i=\braket{\psi_i|\alpha}$.
Moreover, as a consequence of Theorem \ref{theo_1}, $a_{ki}=\braket{\Phi_k|\psi_i}$ are real.
So we have,
\begin{equation}\label{eq:eval_eq}
\sum_{i \in T}\sum_{k} a_{kj} \cot(\frac{\alpha-\phi_k}{2}) a_{ki}~x_i = 0 ~.
\end{equation}
Using the fact that all the $a_{0i}$'s are the same, and equal to $\frac{1}{\sqrt{N}}$, we expand Eq.\eqref{eq:eval_eq} as
\begin{equation}
\frac{1}{N} \cot(\frac{\alpha}{2}) \sum_{i \in T} x_i + \sum_{k>0} \sum_{i \in T} \left( a_{kj} a_{ki} \cot(\frac{\alpha-\phi_k}{2}) + a_{kj} a_{ki} \cot(\frac{\alpha+\phi_k}{2}) \right) x_i = 0 ~.
\end{equation}
With the identity,
$\cot(\frac{\alpha-\phi_k}{2}) + \cot(\frac{\alpha+\phi_k}{2}) = 2\frac{\sin\alpha}{\cos\phi_k-\cos\alpha}$,
it simplifies to
\begin{equation}
\label{eq:mult_master}
\frac{1}{N} \cot(\frac{\alpha}{2}) \sum_{i \in T} x_i = 2 \sum_{k>0}\sum_{i \in T} \left( \frac{a_{kj} a_{ki}\sin\alpha}{\cos\alpha-\cos\phi_k} \right) x_i ~.
\end{equation}
This system of $M$ independent equations, each corresponding to one value of $j$, is the main tool we use to analyse our search algorithm.
It is a multiple target generalization of the corresponding equation in Refs.~\cite{ambainis2005coins,tulsi2012general}, where quantum spatial search with only one target state was studied.
The price we pay for the generalization is the introduction of $M$ unknown coefficients $x_i$.

We combine these $M$ equations in two different ways to obtain identities that are useful in derivation of bounds for spatial search in the next section.
Adding all the equations, we obtain the result
\begin{equation}\label{eq:mult_eval0}
\frac{M}{N}~\text{cosec}^2\frac{\alpha}{2}~(\sum_{i \in T}x_i) = 4 \sum_{k>0} \frac{(\sum_{j \in T}a_{kj}) (\sum_{i \in T}a_{ki}x_i)}{\cos\alpha-\cos\phi_k} ~.
\end{equation}
Furthermore, adding all the equations with weights $x_j^*$, we obtain a different result that is symmetric in $i$ and $j$,
\begin{equation}\label{eq:mult_eval}
\frac{1}{N}~\text{cosec}^2\frac{\alpha}{2}~(\sum_{j \in T}x_j^*) (\sum_{i \in T}x_i) = 4 \sum_{k>0} \frac{(\sum_{j \in T}a_{kj}x_j^*) (\sum_{i \in T}a_{ki}x_i)}{\cos\alpha-\cos\phi_k} ~.
\end{equation}

We cast Eq.\eqref{eq:eval_eq} in another form, by defining an $M \times M$ matrix $B(\alpha)$ with entries:
\begin{equation}\label{eq:B_def}
B(\alpha)_{ij} = \sum_{k}a_{kj} \cot(\frac{\alpha-\phi_k}{2}) a_{ki} ~.
\end{equation} 
Let $\vec{x}$ be the vector of $x_i$ coefficients. Then
\begin{equation}
\label{eq:Bxzero}
B(\alpha) \vec{x} = 0 ~.
\end{equation}  
This means that for the eigenphase $\alpha$ of $U$, $B(\alpha)$ is singular.%
\footnote{In case of a single target vertex, $B(\alpha)=0$ \cite{ambainis2005coins,tulsi2012general}.} 
~$B(\alpha)$ is a real symmetric matrix, so there exists some orthogonal matrix $V$ such that $V B V^T$ is diagonal.
Choosing the first eigenvalue of $B$ to be zero, $V_{0j}=x_j$, and the first row and column of $V B V^T$ vanish:
\begin{equation}
\sum_{i,j \in T} V_{i^\prime i} B(\alpha)_{ij} V_{0j} = 0 ~, \quad \forall i^\prime \in T ~.
\end{equation}

Now we use this alternate form of Eq.\eqref{eq:eval_eq} to find an expression for $\ket{\alpha}$. 
We define a set of $M$ vectors, for $j \in T$,
\begin{equation} 
\label{eq:wjdef}
\ket{w^\alpha_j } = \sum_k a_{kj} \cot(\frac{\alpha-\phi_k}{2}) \ket{\Phi_k} ~.
\end{equation} 
Since $\ket{\psi_i} = \sum_k a_{ki} \ket{\Phi_k}$, they are connected to $B(\alpha)$ as
\begin{equation}
\label{eq:Bdef}
\braket{w^\alpha_j | \psi_i} =  B(\alpha)_{ij} ~,
\end{equation} 
and
\begin{equation}
\sum_{i,j \in T} V_{i^\prime i} ~\braket{w^\alpha_j | \psi_i}~ V_{0j} = 0 ~, \quad \forall i^\prime \in T ~.
\end{equation}
Here the vectors $\sum_i V_{i^\prime i}\ket{\psi_i}$, for $i^\prime \in T$, fully span the reflecting subspace of the oracle $O$ (i.e. the subspace corresponding to $T$).
So the vector $\sum_{j \in T} V_{0j} \ket{w^\alpha_j} = \sum_{j \in T} x_j \ket{w^\alpha_j}$ has no component in this subspace, implying that it is an eigenvector of $O$ with eigenvalue $1$.
Using these properties, we now explicitly construct the eigenvector of $U$ corresponding to the smallest eigenphase.

\smallskip
\begin{lemma}
\label{lemma:0}
$\frac{1}{\mathcal{N}}\sum_{i \in T} x_i (\ket{\psi_i} + i\ket{w^\alpha_i})$ is the eigenvector of $U$ with eigenvalue $e^{i\alpha}$.
Here $\mathcal{N}$ is the normalization factor.
\end{lemma}

\noindent{\bf Proof:} 
By straightforward evaluation, with $U=WO$,
\begin{align}
U \sum_{i \in T} x_i (\ket{\psi_ i} + i \ket{w^\alpha  _i}) &= W  \sum_{i,k} x_i~a_{ki} (-1 + i \cot(\frac{\alpha - \phi_k}{2})) \ket{\Phi _k},\\ 
&= \sum_{i,k} x_i~a_{ki} ~e^{i \phi_k} (-1 + i\cot(\frac{\alpha - \phi_k}{2})) \ket{\Phi _k},\\
&= e^{i\alpha}\sum_{i,k}  x_i~a_{ki}  (1 + i\cot(\frac{\alpha - \phi_k}{2})) \ket{\Phi _k},\\
&= e^{i\alpha} \sum_{i \in T} x_i (\ket{\psi_i} + i \ket{w^\alpha_i}).
\end{align}
Here the third line follows from the identity
$$e^{i\phi} (-1 + i\cot\frac{\alpha-\phi}{2}) =  e^{i\phi} \frac{ie^{i(\alpha-\phi)/2}}{\sin\frac{\alpha-\phi}{2}} = e^{i\alpha} \frac{ie^{-i(\alpha-\phi)/2}}{\sin\frac{\alpha-\phi}{2}} = e^{i\alpha} (1 + i\cot\frac{\alpha-\phi}{2}) ~.
\qquad\qquad\square$$ 

\subsection{Tulsi's controlled spatial search technique}\label{sec:Tulsi}

Tulsi proposed a method to make the spatial search algorithm faster, where the quantum walk and oracle are controlled using an ancilla qubit \cite{tulsi2008faster}.
It is particularly useful when the spectral gap is small, and can be easily implemented as illustrated in Appendix B.
It can be viewed as spatial search on an expanded graph, with a trap vertex attached to every target vertex, as follows:\\
(i) On the expanded $N+M$ vertex graph, the walk operator has the block-diagonal form,
\begin{equation}
\tilde{W} = \begin{pmatrix}
            W & 0\cr 0 & -\mathbb{I}\cr
            \end{pmatrix} ~,
\end{equation}
which corresponds to $M$ additional $\ket{\Phi_{k=t \in T}}$ modes with $\phi_k=\pi$.\\
(ii) The search oracle is $O_\delta = \mathbb{I}-2P_\delta$, with the projection operator
\begin{equation}
P_\delta = \sum_{i \in T} \ket{\psi_i}\ket{\delta}\bra{\psi_i}\bra{\delta} \equiv \sum_{i \in T} \ket{\psi_{i,\delta}}\bra{\psi_{i,\delta}} ~.
\end{equation}
Here the ancilla state is $\ket{\delta}=\cos\delta\ket{0}+\sin\delta\ket{1}$, where $\ket{0}$ labels the marked vertices on the original graph and $\ket{1}$ labels the trap vertices on the expanded graph.\\
(iii) The search operator is $U_\delta=\tilde{W} O_\delta$, and the control parameter $\delta$ is tuned to optimize the search process.
$\delta=0$ gives back the original spatial search algorithm.

We label the states and the operators for search on the expanded graph with the subscript $\delta$.
All our results following from Eq.\eqref{eq:eval_eq} extend to this setting, once we expand the complete set of states $\{\ket{\Phi_k}\}$ to include the states $\ket{\Phi_t}$ and replace $\ket{\psi_i}$ by $\ket{\psi_{i,\delta}}$.
For brevity, we use the notation
\begin{equation}
\ket{\tilde{\Phi}_0}\equiv\ket{\Phi_0}\ket{0} ~,~~
\ket{\tilde{\Phi}_k}\equiv\ket{\Phi_k}\ket{0} ~,~~
\ket{\Phi_{k=t \in T}}\equiv\ket{\psi_t}\ket{1} ~.
\end{equation}
These changes give
\begin{equation}
\braket{\tilde{\Phi}_k|\psi_{i,\delta}} = a_{ki}\cos\delta ~,~~
\braket{\Phi_t|\psi_{i,\delta}} = \delta_{ti}\sin\delta ~.
\end{equation}
Then separation of $k=0$ and $k=t$ contributions modifies Eq.\eqref{eq:mult_master} to
\begin{equation}
\label{eq:mult_master_tulsi}
\frac{1}{N} \cot(\frac{\alpha_\delta}{2}) \sum_{i \in T} x_{i,\delta} - \tan^2\delta \tan(\frac{\alpha_\delta}{2})~x_{j,\delta} = 2 \sum_{k>0}\sum_{i \in T} \left( \frac{a_{kj} a_{ki}\sin\alpha_\delta}{\cos\alpha_\delta-\cos\phi_k} \right) x_{i,\delta} ~,
\end{equation}
and their combined identities, Eqs.\eqref{eq:mult_eval0} and \eqref{eq:mult_eval}, become
\begin{equation}\label{eq:mult_eval0_tulsi}
(\frac{M}{N}~\text{cosec}^2\frac{\alpha_\delta}{2} - \tan^2\delta \sec^2\frac{\alpha_\delta}{2}) \sum_{j \in T}x_{j,\delta} = 4 \sum_{k>0} \frac{(\sum_{j \in T}a_{kj}) (\sum_{i \in T}a_{ki}x_{i,\delta})}{\cos\alpha_\delta-\cos\phi_k} ~.
\end{equation}
\begin{equation}\label{eq:mult_eval_tulsi}
\frac{1}{N}~\text{cosec}^2\frac{\alpha_\delta}{2}~|\sum_{i \in T}x_{i,\delta}|^2 - \tan^2\delta \sec^2\frac{\alpha_\delta}{2}~(\sum_{i \in T}|x_{i,\delta}|^2) = 4 \sum_{k>0} \frac{|\sum_{i \in T}a_{ki}x_{i,\delta}|^2}{\cos\alpha_\delta-\cos\phi_k} ~.
\end{equation}
The property that all the terms in the preceding equation are positive implies that $\alpha_\delta$ decreases as $\delta$ increases from zero.

We also have, as modification of Eq.\eqref{eq:Bxzero},
\begin{equation}
B_\delta(\alpha_\delta)~\vec{x}_\delta \equiv \left( \cos^2\delta B(\alpha_\delta) - \sin^2\delta\tan(\frac{\alpha_\delta}{2})\mathbb{I} \right) \vec{x}_\delta = 0 ~,
\end{equation}
which keeps $B_{\delta}(\alpha_\delta)$ a singular and real symmetric matrix.
Reality of $B_{\delta}(\alpha_\delta)$ implies that all $x_{i,\delta}$ have the same complex phase, and we choose all $x_{i,\delta}$ to be imaginary in concurrence with Corollary \ref{cor:2}.
Furthermore, Eq.\eqref{eq:wjdef} is modified as, 
\begin{equation}
\label{eq:w_alpha_delta}
\ket{w^\alpha_{j,\delta}} = \cos\delta\ket{w^\alpha_j}\ket{0} - \sin\delta\tan\frac{\alpha_\delta}{2}\ket{\psi_j}\ket{1} ~,
\end{equation}
so that $\braket{w^\alpha_{j,\delta}|\psi_{i,\delta}} = B_\delta(\alpha_\delta)_{ij}$, and $\sum_{j \in T} x_{j,\delta} \ket{w^\alpha_{j,\delta}}$ is an eigenvector of $O_\delta$ with eigenvalue $1$.
Then the lowest eigenvector $\ket{\alpha_\delta}$ of $U_\delta$ has the same decomposition as in Lemma \ref{lemma:0}:
\begin{equation}
\label{eq:alpha_delta}
\ket{\alpha_\delta} = \frac{1}{\mathcal{N}} \sum_{i \in T} x_{i,\delta} (\ket{\psi_{i,\delta}} + i \ket{w^\alpha_{i,\delta}}) ~.
\end{equation}

\section{Analysis of Spatial Search}
\label{sec:QGSanalysis}

In this section, we extend the abstract search framework of Ref.~\cite{ambainis2005coins} to the case of multiple target vertices.
The abstract search tries to find one of the target vertices $v \in V$, starting from the easily prepared state $\ket{\psi_s} = \ket{\Phi_0}$.
The algorithm iteratively applies the search operator $U=WO$, $Q$ times to the state $\ket{\psi_s}$.
For the algorithm to succeed, $U^Q\ket{\psi_s}$ must reach a significant overlap with the target subspace spanned by the set $\{\ket{\psi_{i}} ~|~ i \in T\}$. 
We find that this method doesn't succeed with high probability for the flip-flop quantum walk, and gives a subpar oracle complexity as a result (the details are provided in Appendix D).
So, in this section, we use the improved operators with a tunable parameter $\delta$, as per Tulsi's controlled spatial search technique described in Section \ref{sec:Tulsi}.
We find that, with $\tan\delta=\Theta(\frac{1}{\sqrt{g}})$, the algorithm succeeds with $\Theta(1)$ probability without any need for amplitude amplification.
To this end, we prove three technical lemmas, similar to the ones in Ref.~\cite{ambainis2005coins}, to analyse and to bound the runtime and success probability of our search algorithm.
 
The analysis of the search algorithm is simplified by finding the subspace $\tilde{\mathcal{H}} \subset \mathbb{C}^{2Nd}$, left invariant by the operator $U_\delta$.
As shown in Appendix C, for non-bipartite graphs, the $2N+M-1$ dimensional subspace spanned by the eigenvectors corresponding to the non-real eigenvalues of $\tilde{W}$, $\ket{\Phi_0}\ket{0}$ and $\ket{\psi_{i \in T}}\ket{1}$ is the invariant subspace.
For bipartite graphs, the $2N+M-2$ dimensional subspace spanned by the eigenvectors corresponding to the non-real eigenvalues of $\tilde{W}$, $\ket{\Phi_0}\ket{0}$, $\ket{\Phi_b}\ket{0}$ and $\ket{\psi_{i \in T}}\ket{1}$ is the invariant subspace.
 
In the invariant subspace $\tilde{\mathcal{H}}$, the following conditions are sufficient for the algorithm to succeed \cite{ambainis2005coins}, and they are clearly satisfied by the flip-flop quantum search operator:\\
$\bullet$ $\tilde{W}$ has only one eigenvector with eigenvalue $1$ in $\mathcal{H}$, and that eigenvector is $\ket{\psi_s}$.\\
$\bullet$ $\tilde{W}$ is a real matrix.\\
$\bullet$ The search algorithm uses an oracle of the form $O_\delta = \mathbb{I} - 2\sum_{i \in T}\ket{\psi_{i,\delta}}\bra{\psi_{i,\delta}}$.

The first lemma derives asymptotic bounds for the smallest eigenphase $\alpha_\delta$ of $U_\delta$.
In a slight abuse of notation (justified by Theorem \ref{theo_1}), we choose the eigenvalues of the adjacency matrix as $1 = \cos\phi_0 > \cos\phi_1 \ge \ldots \ge \cos\phi_{N-1} \ge -1$.
The spectral gap of $A$ is
\begin{equation}
g = \cos\phi_0 - \cos\phi_1 = 1 - \cos\phi_1 = 2\sin^2(\phi_1/2) ~.
\end{equation}
As before, $\ket{\Phi_k}$ is the eigenvector of $W$ corresponding to the eigenvalue $e^{i\phi_k}$, and by extension $\ket{\Phi_k^*}$ corresponds to the eigenvalue $e^{-i\phi_k}$. 

Now we determine the scaling of $\alpha_\delta$, when $\delta$ is tuned to an optimal value that guarantees $\Theta(1)$ success probability for the search algorithm.

\smallskip
\begin{lemma}
\label{lemma:1}
Let $e^{i\alpha_\delta}$ be the eigenvalue of $U_\delta$ closest to $1$.
Then in terms of the spectral gap of the adjacency matrix $g$, the number of graph vertices $N$, and the number of target vertices $M$, we have $\alpha_\delta = \Theta(\sqrt{\frac{gM}{N}})$, when $\tan{\delta} = \Theta(\frac{1}{\sqrt{g}})$.
\end{lemma}

\noindent{\bf Proof:} 
Here we assume that $\alpha_\delta<\phi_1/2$.
In the case $\alpha_\delta\ge\phi_1/2$, we automatically have $\alpha_\delta = \Theta(\sqrt{g})$, which is stronger than what we want to prove.
With $0<\cos\alpha_\delta-\cos\phi_k<2$, Eq.\eqref{eq:mult_eval_tulsi} gives
\begin{equation}\label{eq:mult_eval_0}
\frac{1}{N}~\text{cosec}^2\frac{\alpha_\delta}{2} |\sum_{i \in T} x_{i,\delta}|^2 - \tan^2\delta\sec^2\frac{\alpha_\delta}{2} (\sum_{i \in T} |x_{i,\delta}|^2) > 2 \sum_{k>0} |\sum_{i \in T} a_{ki} x_{i,\delta}|^2 ~.
\end{equation}
Here the left hand side is positive, because the right hand side of Eq.\eqref{eq:mult_eval_tulsi} is positive.
Using the completeness relation, $\sum_{k>0} a_{ki}a_{kj} = \delta_{ij}- \frac{1}{N}$, we get
\begin{equation}
\frac{R^2}{N}~\text{cosec}^2\frac{\alpha_\delta}{2} - \tan^2\delta\sec^2\frac{\alpha_\delta}{2} > 2 (1-\frac{R^2}{N}) ~,
\end{equation}
where $R^2 = \frac{|\sum_{i \in T} x_{i,\delta}|^2}{\sum_{i \in T} |x_{i,\delta}|^2} \in [1,M]$.
This expression yields the bound,
\begin{equation}\label{eq:R_bound}
R^2 > \frac{N}{1+2\sin^2(\alpha_\delta/2)} \tan^2{\delta}\tan^2\frac{\alpha_\delta}{2} ~,
\end{equation}
which we use to obtain a lower bound on $\alpha_\delta$.

Next, we express the combinations on the right hand side of Eq.\eqref{eq:mult_eval0_tulsi} as the matrix elements of $P = \sum_{i \in T} \ket{\psi_i}\bra{\psi_i}$:   
\begin{equation}
\sum_{j \in T} a_{kj} = \sum_{j \in T} \braket{\Phi_k|\psi_j}
  = \sqrt{N} \sum_{j \in T} \braket{\Phi_k|\psi_j} \braket{\psi_j|\Phi_0 }
  = \sqrt{N} \braket{\Phi_k|P|\Phi_0} ~,
\end{equation}
\begin{equation}
\sum_{i \in T} a_{ki}x_{i,\delta}
  = \sum_{i \in T} \braket{\Phi_k|\psi_i} \braket{\psi_i|X_\delta}
  = \braket{\Phi_k|P|X_\delta} ~.
\label{eq:A_alpha}
\end{equation}
Here $\ket{X_\delta}$ is a vector such that $\braket{\psi_{i,\delta}|X_\delta} = x_{i,\delta}$.
Then Eq.\eqref{eq:mult_eval0_tulsi} can be written as
\begin{equation}
(\frac{M}{N}~\text{cosec}^2\frac{\alpha_\delta}{2} - \tan^2\delta\sec^2\frac{\alpha_\delta}{2}) \sum_{i \in T} x_{i,\delta} = 2\sqrt{N} \sum_{k\ne0} \frac{\braket{\Phi_0|P|\Phi_k} \braket{\Phi_k|P|X_\delta}} {\cos\alpha_\delta-\cos\phi_k} ~. 
\end{equation}
Defining the operator
\begin{equation}
\label{eq:V_alpha}
V_{\alpha_\delta} = \sum_{k\ne0} \frac{\ket{\Phi_k}\bra{\Phi_k}}{\cos\alpha_\delta-\cos\phi_k}
~,
\end{equation}
we convert the right hand side of the preceding equation to a matrix element,
\begin{equation}
(\frac{M}{N}~\text{cosec}^2\frac{\alpha_\delta}{2} - \tan^2\delta\sec^2\frac{\alpha_\delta}{2}) \sum_{i \in T} x_{i,\delta} = 2\sqrt{N}\braket{\Phi_0|P V_{\alpha_\delta} P|X_\delta} ~.
\end{equation}
Matrix elements satisfy, $|\braket{x,Ay}| \le \|x\|_2~\|A\|_s~\|y\|_2$, where $\|A\|_s$ is the spectral norm of $A$.
So taking absolute values, we get
\begin{equation}\label{eq:matrix_element}
|\frac{M}{N} ~ \text{cosec}^2\frac{\alpha_\delta}{2} - \tan^2\delta\sec^2\frac{\alpha_\delta}{2}|~|\sum_{i \in T} x_{i,\delta}| \le 2\sqrt{N}~\|P\ket{\Phi_0}\|_2~\|V_{\alpha_\delta}\|_s~\|P\ket{X_\delta}\|_2 ~.
\end{equation} 
From the definition of $V_{\alpha_\delta}$, and the assumption that $\alpha_\delta<\phi_1/2$, we see that its spectral norm is $\frac{1}{\cos\alpha_\delta-\cos\phi_1}$.
Also, $\|P\ket{\Phi_0}\|_2 = \sqrt{M/N}$, and $\|P\ket{X_\delta}\|_2 = \sqrt{\sum_{i \in T} |x_{i,\delta}|^2}$.
Therefore,
\begin{equation}
R~|\frac{M}{N} ~ \text{cosec}^2\frac{\alpha_\delta}{2} - \tan^2\delta\sec^2\frac{\alpha_\delta}{2}| \le \frac{2\sqrt{M}}{\cos\alpha_\delta-\cos\phi_1} ~.
\end{equation} 

Now to determine the scaling of $\alpha_\delta$ with $g$, we observe that for $\alpha_\delta<\phi_1/2$,
\begin{equation}
\cos\alpha_\delta-\cos\phi_1 > \cos\frac{\phi_1}{2}-\cos\phi_1 > \cos^2\frac{\phi_1}{2}-\cos\phi_1 = \frac{1-\cos\phi_1}{2} = \frac{g}{2} ~.
\label{eq:g_bound}
\end{equation} 
Also, $\frac{M}{N}\text{cosec}^2\frac{\alpha_\delta}{2} - \tan^2\delta\sec^2\frac{\alpha_\delta}{2} > 0$, since the left hand side of Eq.\eqref{eq:mult_eval_0} is positive and $R^2\le M$.
So we have,
\begin{equation}\label{eq:mult_eval_1}
0 < \frac{M}{N} ~ \text{cosec}^2\frac{\alpha_\delta}{2} - \tan^2\delta\sec^2\frac{\alpha_\delta}{2} < \frac{4\sqrt{M}}{gR} ~.
\end{equation}

The left inequality in Eq.\eqref{eq:mult_eval_1} gives an upper bound on $\alpha_\delta$, as
\begin{equation}
\tan^2\frac{\alpha_\delta}{2} < \frac{M}{N\tan^2\delta} ~.
\end{equation}
Together with $x<\tan x$ for $x\in[0,\frac{\pi}{2}]$, it becomes
\begin{equation}\label{eq:alpha_upper}
\alpha_\delta < \sqrt\frac{M}{N}\frac{2}{\tan\delta} ~.
\end{equation}
To optimize the search algorithm, we want $\alpha_\delta$ to be as large as possible, and hence this bound limits the values of $\delta$ that we can use.

The right inequality in Eq.\eqref{eq:mult_eval_1} can be reexpressed as, using Eq.\eqref{eq:R_bound},
\begin{align}
\text{cosec}^2\frac{\alpha_\delta}{2} &< \frac{4N}{\sqrt{M}gR} + \frac{N}{M}\tan^2\delta\sec^2\frac{\alpha_\delta}{2} ~,\\ 
&< \sqrt{\frac{N}{M}}\frac{4}{g\tan\delta\tan(\alpha_\delta/2)}\sqrt{1+2\sin^2\frac{\alpha_\delta}{2}} + \frac{N}{M}\tan^2\delta\sec^2\frac{\alpha_\delta}{2} ~.
\end{align}
It can be rearranged as
\begin{equation}
\frac{1}{\sin\alpha_\delta} < \frac{2}{g\tan\delta}\sqrt{\frac{N}{M}(2-\cos\alpha_\delta)} + \frac{N}{2M}\tan^2\delta\tan\frac{\alpha_\delta}{2}\sec^2\frac{\alpha_\delta}{2} ~.
\end{equation}
For $\tan^2\delta = \Theta(\frac{1}{g})$, Eq.\eqref{eq:alpha_upper} implies $\tan\frac{\alpha_\delta}{2}\sec^2\frac{\alpha_\delta}{2} =  \Theta(\alpha_\delta) = O(\sqrt{\frac{gM}{N}})$.
Both the terms on the right hand side then scale as $\Theta(\sqrt{\frac{N}{gM}})$, and using $x>\sin x$ for $x\in[0,\frac{\pi}{2}]$, we get
\begin{equation}\label{eq:alpha_lower}
\alpha_\delta = \Omega(\sqrt{\frac{gM}{N}}).
\end{equation} 
Note that it is the ability to tune $\delta$ in Tulsi's algorithm that has brought together the two bounds, Eq.\eqref{eq:alpha_upper} and \eqref{eq:alpha_lower}, and we have $\alpha_\delta = \Theta(\sqrt{\frac{gM}{N}})$ as stated in the lemma.
\hfill $\square\,$ 

\smallskip
In case of Grover search, the two-dimensional subspace, spanned by the two eigenvectors of the search operator corresponding to the smallest eigenphases $\pm\alpha$, is the invariant subspace of $U$ \cite{nielsen2002quantum}.
That makes the analysis of Grover search extremely simple.
The algorithm is already optimal, and a speed-up by Tulsi's method is not needed.
The situation is more complicated in our QGS algorithm, but we show that the algorithm can be kept largely within the two-dimensional subspace corresponding to the eigenphases $\pm\alpha_\delta$, although not completely within it.
In the following lemma, we prove that the starting state has a sufficiently large overlap with the two-dimensional subspace formed by $\ket{\pm\alpha_\delta}$.

\smallskip
\begin{lemma}
\label{lemma:2}
Let $\ket{w_s} = \frac{\ket{\alpha_\delta}+\ket{-\alpha_\delta}}{\sqrt{2}}$.
Then, provided $\alpha_\delta<\phi_1/2$,
\begin{equation}
\frac{1}{|\braket{\tilde{\Phi}_0|w_s}|^2} < 1 + \frac{\alpha_\delta^2}{g} ~.
\end{equation}
\end{lemma}   
   
\noindent{\bf Proof:} 
Since $U_\delta$ is real, we have $\ket{-\alpha_\delta}=\ket{\alpha_\delta}^*$.
Moreover, we have chosen $x_{i,\delta}$ to be purely imaginary.
So Eqs.\eqref{eq:alpha_delta} and \eqref{eq:w_alpha_delta} give
\begin{equation}
\ket{w_s} = \frac{i}{\sqrt{2}\mathcal{N}} \sum_{i \in T} x_{i,\delta} \left( \cos\delta(\ket{w_i^\alpha}+\ket{w_i^\alpha}^*)\ket{0} - 2\sin\delta \tan\frac{\alpha_\delta}{2}\ket{\psi_i}\ket{1} \right) ~.
\end{equation}
Now, using the definition of Eq.\eqref{eq:wjdef}, and noting the fact that $\ket{\Phi_k}^*$ has the eigenphase $-\phi_k$ can be used to flip the label $k$,
\begin{align}
\ket{w_s} &= \frac{i}{\sqrt{2}\mathcal{N}} \sum_{i \in T} x_{i,\delta} \left( \cos\delta \sum_k a_{ki}(\cot\frac{\alpha_\delta-\phi_k}{2} + \cot\frac{\alpha_\delta+\phi_k}{2})\ket{\Phi_k}\ket{0} - 2\sin\delta\tan\frac{\alpha_\delta}{2}\ket{\psi_i}\ket{1} \right) ,\nonumber\\
          &= -\frac{i\sqrt{2}}{\mathcal{N}} \sum_{i \in T} x_{i,\delta} \left( \cos\delta \sum_k a_{ki}(\frac{\sin\alpha_\delta}{\cos\alpha_\delta-\cos\phi_k})\ket{\Phi_k}\ket{0} - \sin\delta\tan\frac{\alpha_\delta}{2}\ket{\psi_i}\ket{1} \right) .
\end{align}
Separating the $k=0$ contribution, we get
\begin{align}
\ket{w_s} &= \frac{i\sqrt{2}\cos\delta}{\mathcal{N}\sqrt{N}} \cot\frac{\alpha_\delta}{2} (\sum_{i \in T} x_{i,\delta}) \ket{\Phi_0}\ket{0} - \frac{i\sqrt{2}\cos\delta}{\mathcal{N}} \sum_{k\ne0} (\sum_{i \in T} a_{ki}x_{i,\delta}) (\frac{\sin\alpha_\delta}{\cos\alpha_\delta-\cos\phi_k}) \ket{\Phi_k}\ket{0} \nonumber\\ 
          & - \frac{i\sqrt{2}\sin\delta}{\mathcal{N}} \tan\frac{\alpha_\delta}{2} \sum_{i \in T} x_{i,\delta} \ket{\psi_i}\ket{1} ~.
\end{align}
which gives the overlap,
\begin{equation}
\label{eq:inner_prod}
D_s \equiv |\braket{\tilde{\Phi}_0|w_s}|^2 = \frac{2\cos^2\delta}{\mathcal{N}^2 N} \cot^2\frac{\alpha_\delta}{2} |\sum_{i \in T} x_{i,\delta}|^2 ~.
\end{equation}

To show that this overlap is large enough we need to evaluate $\mathcal{N}$.
$\ket{w_s}$ is a unit vector, because $\ket{\alpha_\delta}$ and $\ket{-\alpha_\delta}$ are orthogonal.
That allows us to evaluate $\mathcal{N}$, as
\begin{align}
\label{eq:norm_factor1}
\mathcal{N}^2 &= \frac{2\cos^2\delta}{N} \cot^2\frac{\alpha_\delta}{2} |\sum_{i \in T} x_{i,\delta}|^2 + 4\cos^2\delta \sum_{k>0} |\sum_{i \in T}a_{ki}x_{i,\delta}|^2 \frac{\sin^2\alpha_\delta}{(\cos\alpha_\delta-\cos\phi_k)^2} \nonumber\\
              &+ 2\sin^2\delta\tan^2\frac{\alpha_\delta}{2} \sum_{i \in T} |x_{i,\delta}|^2 ~.
\end{align}
Combining this expression with Eq.\eqref{eq:inner_prod}, we get
\begin{align}
\frac{1}{D_s} &= 1 + \frac{\sum_{k>0} |\sum_{i \in T} a_{ki}x_{i,\delta}|^2 \frac{2N\sin^2\alpha_\delta}{(\cos\alpha_\delta-\cos\phi_k)^2} } {\cot^2(\alpha_\delta/2) |\sum_{i \in T} x_{i,\delta}|^2} + N\tan^2\delta \frac{ \tan^4(\alpha_\delta/2) \sum_{i \in T} |x_{i,\delta}|^2}{|\sum_{i \in T} x_{i,\delta}|^2} ~,\nonumber\\
\begin{split}\label{eq:ws_bound}
  &< 1 + \frac{8N\sin^4(\alpha_\delta/2)}{|\sum_{i \in T}x_{i,\delta}|^2 (\cos\alpha_\delta-\cos\phi_1)} \sum_{k>0} \frac{|\sum_{i \in T}a_{ki}x_{i,\delta}|^2}{\cos\alpha_\delta-\cos\phi_k} \\
  &\qquad + N\tan^2\delta \frac{\tan^4(\alpha_\delta/2) \sum_{i \in T} |x_{i,\delta}|^2}{|\sum_{i \in T} x_{i,\delta}|^2} ~.
\end{split}
\end{align} 
where, in the last step, we have used $(\cos\alpha_\delta-\cos\phi_k) > (\cos\alpha_\delta-\cos\phi_1)$ for $\alpha_\delta<\phi_1$.
Further simplifying this result using Eq.\eqref{eq:mult_eval_tulsi}, we obtain
\begin{equation}\label{eq:ws_bound_1}
\frac{1}{D_s} < 1 + \frac{2\sin^2(\alpha_\delta/2)}{\cos\alpha_\delta-\cos\phi_1} - N\tan^2\delta \frac{\tan^4(\alpha_\delta/2) \sum_{i \in T} |x_{i,\delta}|^2}{|\sum_{i \in T} x_{i,\delta}|^2} (\frac{1+\cos\phi_1}{\cos\alpha_\delta-\cos\phi_1}) ~.
\end{equation}
Now, Eq.\eqref{eq:g_bound} holds when $\alpha_\delta<\phi_1/2$, and the last term on the right hand side is negative, which gives the bound stated in the lemma,
\begin{equation}
\frac{1}{D_s} < 1 + \frac{\alpha_\delta^2}{g} ~.
\end{equation}
We note that this bound is finite for any value of the parameter $\delta$, and the overlap $D_s$ increases as $\delta$ increases from zero.
\hfill $\square\,$   

\smallskip
The next lemma quantifies the success probability of our QGS algorithm, by showing that the final state also has a sufficiently large overlap with the two-dimensional subspace formed by $\ket{\pm\alpha_\delta}$.

\smallskip
\begin{lemma}
\label{lemma:3}
Let $\ket{w_t} = \frac{\ket{\alpha_\delta}-\ket{-\alpha_\delta}}{\sqrt{2}}$.
Then, provided $\alpha_\delta<\phi_1/2$ and $\tan\delta = \frac{1}{\sqrt{g}}$,
\begin{equation}
\|P_\delta\ket{w_t}\|  =  \Omega(1) ~.
\end{equation}  
\end{lemma}   

\noindent{\bf Proof:} 
We have noted that $P_\delta\ket{w^\alpha_{i,\delta}}=0$, and have chosen $x_{i,\delta}$ to be purely imaginary.
So Eq.\eqref{eq:alpha_delta} gives,
\begin{equation}
\label{eq:overlap_lemma3}
\|P_\delta\ket{w_t}\|^2 = \|P_\delta \frac{\sqrt{2}}{\mathcal{N}} \sum_{i \in T} x_{i,\delta}\ket{\psi_{i,\delta}}\|^2 = \frac{2}{\mathcal{N}^2} \sum_{i \in T} |x_{i,\delta}|^2 ~.  
\end{equation}
Using the expression for $\mathcal{N}^2$ from  Eq.\eqref{eq:norm_factor1}, and repeating the same steps as in Lemma \ref{lemma:2},
\begin{align}
\label{eq:pwt_bound1}
\frac{1}{\|P_\delta\ket{ w_t}\|^2} &=  \cos^2\delta\cot^2\frac{\alpha_\delta}{2} \frac{|\sum_{i \in T}x_{i,\delta}|^2}{N\sum_{i \in T}|x_{i,\delta}|^2} + 2 \cos^2\delta \sum_{k>0} \frac{\sin^2 \alpha_\delta}{(\cos\alpha_\delta-\cos\phi_k)^2} \frac{|\sum_{i \in T}a_{ki}x_{i,\delta}|^2}{\sum_{i \in T}|x_{i,\delta}|^2} \nonumber\\
&+ \sin^2\delta\tan^2\frac{\alpha_\delta}{2} ~,\\
\label{eq:pwt_bound2}
&< \cos^2\delta\cot^2\frac{\alpha_\delta}{2} \frac{|\sum_{i \in T}x_{i,\delta}|^2}{N\sum_{i \in T}|x_{i,\delta}|^2} + \frac{2\cos^2\delta\sin^2\alpha_\delta}{\cos\alpha_\delta-\cos\phi_1} \sum_{k>0} \frac{|\sum_{i \in T}a_{ki}x_{i,\delta}|^2}{(\cos\alpha_\delta-\cos\phi_k)\sum_{i \in T}|x_{i,\delta}|^2} \nonumber\\
& + \sin^2\delta\tan^2\frac{\alpha_\delta}{2} ~,\\
\label{eq:pwt_bound3}
&= \cos^2\delta\cot^2\frac{\alpha_\delta}{2} \frac{|\sum_{i \in T}x_{i,\delta}|^2}{N\sum_{i \in T}|x_{i,\delta}|^2} (\frac{1-\cos\phi_1}{\cos\alpha_\delta-\cos\phi_1}) - \sin^2\delta\tan^2\frac{\alpha_\delta}{2} (\frac{1+\cos\phi_1}{\cos\alpha_\delta-\cos\phi_1}) .
\end{align}
Here we have used Eq.\eqref{eq:mult_eval_tulsi} in the last step.
This result is further simplified, using $|\sum_{i \in T}x_{i,\delta}|^2 \le M\sum_{i \in T}|x_{i,\delta}|^2$ and Eq.\eqref{eq:g_bound}, to
\begin{equation}\label{eq:pwt_bound4}
\frac{1}{\|P_\delta\ket{w_t}\|^2} < \cos^2\delta\cot^2\frac{\alpha_\delta}{2} (\frac{2M}{N}) - \sin^2\delta\tan^2\frac{\alpha_\delta}{2} (\frac{2-g}{g}) ~.
\end{equation}

Now we enforce the results from Lemma \ref{lemma:1}.
$\tan^2\delta=\Theta(\frac{1}{g})$ implies $\cos^2\delta=\Theta(g)$, and $\alpha_\delta^2=\Theta(\frac{gM}{N})$ implies $\cot^2\frac{\alpha_\delta}{2}=\Theta(\frac{N}{gM})$.
Also, the last term on the right hand side is negative.
As a result,
\begin{equation}
\frac{1}{\|P_\delta\ket{w_t}\|^2} < \Theta(1) ~,
\end{equation}
as stated in the lemma.
Note that for $\delta=0$, we only have $\frac{1}{\|P\ket{w_t}\|^2} < \Theta(\frac{M}{\alpha^2 N}) = \Theta(\frac{1}{g})$.
Thus tuning of the parameter $\delta$ improves the success probability of the algorithm substantially when $g$ is small.%
\footnote{This behaviour is fully consistent with the estimate $\cos^2\delta=\|P\ket{w_t}\|^2$, used in Ref.~\cite{patel2012search} to tune the algorithm in absence of knowledge of $g$.}
\hfill $\square\,$   

\smallskip
The preceding three technical lemmas give us all the necessary ingredients to analyse the performance of the QGS algorithm.
We now show that $Q = \lfloor\frac{\pi}{2\alpha_\delta}\rfloor$ iterations of the search operator on the starting state $\ket{\tilde{\Phi}_0}$ produces a significant overlap with the target subspace.
The scaling of $\alpha_\delta$ given by Lemma \ref{lemma:1} then produces the oracle complexity $\Theta(\sqrt{\frac{N}{gM}})$.
Since $g=O(1)$, and the optimal Grover result is $\Theta(\sqrt{\frac{N}{M}})$, the extra factor of $\frac{1}{\sqrt{g}}$ can be viewed as the price paid for making the quantum walk local.

\begin{theorem}
\label{theo:3}
The QGS algorithm, on regular graphs with $M=o(N)$, succeeds with $\Theta(1)$ probability for oracle complexity $\Theta(\sqrt{\frac{N}{gM}})$, when $\tan\delta=\Theta(\frac{1}{\sqrt{g}})$. 
\end{theorem}

\noindent{\bf Proof:}
The performance of the QGS algorithm is essentially governed by the overlap of the quantum state with the two-dimensional subspace formed by the states $\ket{\alpha_\delta}$ and $\ket{-\alpha_\delta}$.
We have evaluated the overlaps of the starting and the final states of the QGS algorithm, with this two-dimensional subspace, in Lemmas \ref{lemma:2} and \ref{lemma:3} respectively, assuming the condition that $\alpha_\delta<\phi_1/2$.
This condition is satisfied for $M=o(N)$ because $\phi_1=\Theta(\sqrt{g})$, and $\alpha_\delta=\Theta(\sqrt{\frac{gM}{N}})$ for $\tan\delta=\Theta(\frac{1}{\sqrt{g}})$ according to Lemma \ref{lemma:1}.

Now define the residual vector $\ket{\phi_R} = \ket{\tilde{\Phi}_0}-\ket{w_s}$.
From Lemma \ref{lemma:2}, we see that
\begin{equation}
\|\ket{\phi_R}\| = \sqrt{2-2\braket{\tilde{\Phi}_0|w_s}} = O(\frac{\alpha_\delta}{\sqrt{g}}) ~.
\end{equation}   
It follows that $\|U_\delta^Q\ket{\tilde{\Phi}_0} - U_\delta^Q\ket{w_s}\| = O(\frac{\alpha_\delta}{\sqrt{g}})$.
Moreover,
\begin{align}
U_\delta^Q\ket{w_s} &= \frac{e^{i\alpha_\delta Q}\ket{\alpha_\delta} + e^{-i\alpha_\delta Q}\ket{-\alpha_\delta}}{\sqrt{2}} ~,\\
             &= i\ket{w_t}~(1+O(\alpha_\delta^2)) + \ket{w_s}~O(\alpha_\delta) ~,\quad \text{for} ~ Q=\lfloor\frac{\pi}{2\alpha_\delta}\rfloor ~,
\end{align}
implies that%
\footnote{We have used the big $O$ symbol here to denote a vector of that length.}
\begin{equation}
U_\delta^Q\ket{\tilde{\Phi}_0} = i\ket{w_t}~(1+O(\alpha_\delta^2)) + \ket{w_s}~O(\alpha_\delta) + O(\frac{\alpha_\delta}{\sqrt{g}}) ~.
\end{equation}

After $Q$ iterations, the probability of the final state to be in the target subspace is,
\begin{equation}
p_s = \|P_\delta U_\delta^Q\ket{\tilde{\Phi}_0}\|^2 ~.
\end{equation}
We have $\|P_\delta\ket{w_t}\|=\Omega(1)$ as per Lemma \ref{lemma:3}, and $\alpha_\delta = \Theta(\sqrt{\frac{gM}{N}})$ as per Lemma \ref{lemma:1}.
So
\begin{equation}
P_\delta U_\delta^Q\ket{\tilde{\Phi}_0} = iP_\delta\ket{w_t}~(1+O(\alpha_\delta^2)) + O(\sqrt{\frac{M}{N}}) ~,
\end{equation}
and the bound from Lemma \ref{lemma:3} ensures that the magnitude of $P_\delta\ket{w_t}$ dominates over those of the correction terms.

The success probability of the preceding procedure is thus $p_s=\Theta(1)$, and its oracle complexity is $Q = \lfloor\frac{\pi}{2\alpha_\delta}\rfloor = \Theta(\sqrt{\frac{N}{gM}})$.
\hfill $\square\,$ 

\smallskip
This theorem shows that the spectral gap of the graph plays an important role in determining the oracle complexity of the QGS algorithm.
In case of Ramanujan graphs \cite{donetti2006optimal}, $g \ge 1-\frac{2\sqrt{d-1}}{d}$, and our results show that the oracle complexity of the QGS algorithm is optimal in $N$ and $M$, the same as that for Grover search.
If no estimate of $g$ is available, we cannot say anything about the optimality of the algorithm.

In the next two sections, we extend our general analysis to two important examples of graphs, the complete graph and the $D$-dimensional hypercubic lattice, where not only $g$ but the complete spectrum of the adjacency matrix is known.
$g$ is $\Theta(1)$ in the former case, but it is $o(1)$ in the latter case.
Our methods demonstrate that the QGS algorithm achieves optimal scaling of the oracle complexity in both these cases.

\paragraph{Comparison with other spatial search algorithms:}
Spatial search algorithms that deal with multiple targets have been proposed by Magniez et al. \cite{magniez2011walk} and Krovi et al. \cite{krovi2016walk}.
Both these algorithms use the quantum walk framework introduced by Szegedy, and have a wider applicability than our algorithm, but they don't use the quantum walk directly as we have done.
They apply a phase estimation procedure to the quantum walk evolution, and obtain oracle complexity $O(\sqrt{\frac{N}{gM}})$ \cite{dohotaru2017amp}.%
\footnote{The oracle complexity obtained in Ref.~\cite{krovi2016walk} is $\Theta(\sqrt{h_T^+})$, where the extended hitting time obeys $h_T^+<\frac{N}{gM}$.}
~The phase estimation procedure of these algorithms, however, needs larger spatial resources compared to our algorithm.

Explicitly, the algorithm of Magniez et al. requires a fresh set of ancilla qubits at every iteration, which means that it would need $O(\sqrt{\frac{N}{gM}})$ qubits.
In contrast, our algorithm and that of Krovi et al. require only $O(\log N)$ qubits.
The algorithm of Krovi et al. has space complexity worse than ours by a constant factor, due to the ancilla qubits required for phase estimation.
It also uses an interpolation between the random walk and the leaking walk, which is non-trivial to implement.
On the other hand, our algorithm is very easy to implement, it does not involve a phase estimation procedure, and the interpolation implemented by Tulsi's method requires only one extra ancilla qubit. 

Moreover, in Section \ref{sec:QGSlattice}, we will show that our algorithm achieves the optimal oracle complexity of $\Theta(\sqrt{\frac{N}{M}})$, for quantum search on $D>4$ hypercubic lattices with $M=o(N^{\frac{D-2}{D}})$.
It is clearly an improvement over the best known upper bound on the oracle complexity for Krovi et al.'s algorithm for the same problem, $O(\text{min}(\sqrt{\frac{N}{gM}},\sqrt{N}))$ with $g=\Theta(N^{-2/D})$.%
\footnote{The first bound is from Ref.~\cite{dohotaru2017amp}, while the second follows from Theorem 2 and Corollary 3 of Ref.~\cite{hoyer2017multi}.}

We point out that a random $d$-regular graph is a good expander \cite{friedman89random}, with $g=\Theta(1)$.
Our QGS algorithm is therefore asymptotically optimal for almost all regular graphs,%
\footnote{i.e. with probability $p\rightarrow1$, as $N\rightarrow\infty$.}
~and superior to the algorithms of Refs.~\cite{magniez2011walk,krovi2016walk} in terms of simplicity and use of spatial resources.     

\section{Analysis of Spatial Search on the Complete Graph}
\label{sec:QGScomplete}

On the complete graph there are no locality constraints, and we expect the QGS algorithm to perform close to Grover search.
As a matter of fact, the QGS on the complete graph can be solved exactly, which clarifies various connections between the analysis presented in this work and the much simpler analysis of Grover search.
The extra coin degree of freedom makes the Hilbert space for the QGS larger than that for Grover search.
With $d=N-1$ for the complete graph, its Hilbert space of dimension $N(N-1)$ is much larger than the $N$-dimensional Hilbert space of Grover search.
Even the invariant subspace, of dimension $2N-1$ for non-bipartite graphs, is much larger than the $2$-dimensional invariant space of Grover search. 

The adjacency matrix for the complete graph with $N$ vertices has eigenvalues $1$ and $-\frac{1}{N-1}$, where the smaller eigenvalue has multiplicity $N-1$.
The spectral gap is thus $g=\frac{N}{N-1}$, and a speed-up by Tulsi's method is not needed.
The leaking walk matrix, obtained by removing $M$ vertices, is just a scaled version of the adjacency matrix of the complete graph with $N-M$ vertices.
The scale factor is the ratio of the corresponding graph degrees, $\frac{N-M-1}{N-1}$.
So the eigenvalues of $\tilde{A}_T$ are $\frac{N-M-1}{N-1}$ and $-\frac{1}{N-1}$, with multiplicities $1$ and $N-M-1$ respectively.
From Theorem \ref{theo_2}, we then get
\begin{equation}
\cos\alpha = \frac{N-M-1}{N-1} = 1 - \frac{M}{N-1} ~. 
\end{equation}  
This value of $\alpha$ implies that the QGS algorithm requires $Q=\Theta(\sqrt{\frac{N}{M}})$ iterations to succeed.
For comparison, Grover search has $\cos\alpha = 1-\frac{2M}{N}$, corresponding to $\alpha$ being larger by about a factor of $\sqrt{2}$, and the same scaling for $Q$.

Next, the eigenvector corresponding to the largest eigenvalue of $\tilde{A}_T$ is the uniform superposition vector.
So from Eq.\eqref{eq:x_express}, for $i \in T$,
\begin{equation}
x_i = -\frac{i}{(N-1)\sin\alpha} \frac{|N(i) \cap (V-T)|}{\sqrt{N-M}}
    = -\frac{i\sqrt{N-M}}{\sqrt{M(2N-M-2)}} ~,
\label{eq:x_compl_graph}
\end{equation}
which are independent of the vertex label.
The overlap of Eq.\eqref{eq:inner_prod} then becomes, using the identity $\cot\frac{\alpha}{2} = \sqrt{\frac{1+\cos\alpha}{1-\cos\alpha}}$,
\begin{align}
|\braket{\Phi_0|w_s}| &= \frac{\sqrt{2}}{\mathcal{N}\sqrt{N}} ~M|x_i|~ \sqrt{\frac{2N-M-2}{M}} ~,\\
                      &= \frac{\sqrt{2(N-M)}}{\mathcal{N}\sqrt{N}} ~.
\label{eq:grover_overlap_1}
\end{align}
The normalization factor given by Eq.\eqref{eq:norm_factor1} simplifies to:
\begin{align}
\mathcal{N}^2 &= \frac{2M^2}{N}\cot^2(\frac{\alpha}{2})~|x_i|^2 + \frac{4\sin^2\alpha}{(\cos\alpha+\frac{1}{N-1})^2}~|x_i|^2 \sum_{k>0} |\sum_{i \in T} a_{ki}|^2 ~,\\
  &= \frac{2M^2}{N}\cot^2(\frac{\alpha}{2})~|x_i|^2 + \left( \frac{\sin^2\alpha}{\cos\alpha+\frac{1}{N-1}} \right) \frac{M^2}{N}\text{cosec}^2(\frac{\alpha}{2})~|x_i|^2 ~,\\
  &= \left( \frac{2(2N-M-2)}{M} + \frac{2(2N-M-2)}{N-M} \right) \frac{M^2}{N} \frac{N-M}{M(2N-M-2)} ~,\\
  &= 2 ~,
\end{align}
where we have used Eq.\eqref{eq:mult_eval} in the second step. Thus
\begin{equation}
\mathcal{N} = \sqrt{2} ~,~~
|\braket{\Phi_0|w_s}|^2 = 1 - \frac{M}{N} ~.
\end{equation} 
This overlap is only slightly worse than the corresponding value $1$ for Grover search.

For the success probability, Eqs.\eqref{eq:overlap_lemma3} and \eqref{eq:x_compl_graph} produce,
\begin{equation}
\|P\ket{w_t}\|^2 = \frac{2M}{\mathcal{N}^2} |x_i|^2 
                 = \frac{1}{1+\frac{N-2}{N-M}} ~.
\end{equation}
This asymptotically approaches $1/2$, which is worse than Grover search by a factor of $2$.

Overall, the QGS on complete graph is about a factor of $2$ worse in oracle complexity compared to Grover search, and requires about $N$ times larger Hilbert space dimensionality.

\section{Analysis of Spatial Search on the D-dimensional Hypercubic Lattice}
\label{sec:QGSlattice}
 
Now we analyse the QGS algorithm on a $D$-dimensional hypercubic lattice, with nearest neighbor connectivity.
We need $D>2$ in our analysis, which is consistent with tests of the QGS algorithm in numerical simulations \cite{patel2010search}.
The spectral gap of the $D$-dimensional hypercubic lattice is $g_D=\Theta(N^{-2/D})$.
Its direct substitution in the result of Theorem \ref{theo:3} gives an upper bound of $O(N^{\frac{1}{2}+\frac{1}{D}})$ for the oracle complexity of the QGS.
But unlike the general case, the entire spectrum of a $D$-dimensional hypercubic lattice is known; so we improve upon our earlier analysis that focused only on the eigenvalue $\phi_1$, and achieve a better oracle complexity.

\paragraph{Spectrum of the $D$-dimensional lattice:}
A $D$-dimensional hypercubic lattice with periodic boundary conditions is a regular graph of degree $2D$.
We choose $N=L^D$, and the label the vertices of this graph by $D$ integers, $\mathbf{y} = (y_1,\ldots,y_D)$, $y_i \in \{0,1,\ldots,L-1\}$.
The adjacency matrix of this graph is fully diagonalized by a Fourier transform.
Its eigenvalues can be indexed by a discrete momentum vector $\mathbf{k} = (k_1,\ldots,k_D)$, where $k_i \in \{0,1,\ldots,L-1\}$.%
\footnote{It is convenient to use modulo $L$ labels, so that $-k \equiv L-k$.} 
~For each $\mathbf{k}$, the graph has an eigenvalue,
\begin{equation}
\cos\phi_\mathbf{k} = \frac{1}{D} \sum_i \cos(\frac{2\pi k_i}{L}) ~.
\end{equation}
The Fourier coefficients give the $\mathbf{y}$-components of the $\mathbf{k}$-eigenvectors:
\begin{equation}
a_{\mathbf{ky}} = \frac{1}{\sqrt{N}} \exp(\frac{2\pi i\mathbf{k}\cdot\mathbf{y}}{L}) ~.
\end{equation}
These $a_{\mathbf{ky}}$ are complex as per the standard convention, whereas we have assumed them to be real in all our proofs.
That is not a problem because $\cos\phi_\mathbf{k} = \cos\phi_\mathbf{-k}$, and hence both $\mathbf{k}$ and $-\mathbf{k}$ belong to the same two dimensional eigenspace of the adjacency matrix.
In strict adherence to our rules, we should replace $a_{\pm\mathbf{ky}}$ by the real linear combinations $\sqrt{\frac{2}{N}} \cos(\frac{2\pi\mathbf{k}\cdot\mathbf{y}}{L})$ and $\sqrt{\frac{2}{N}} \sin(\frac{2\pi\mathbf{k}\cdot\mathbf{y}}{L})$.
But that would have no effect on our results, provided we take care to replace $a_{ki}^2$ by $|a_{ki}|^2$---ultimately we only deal with the projections of the eigenvectors on various eigenspaces of $A$, which remain unchanged.

As before, to analyse the algorithm, we need lower bounds on $\alpha$, $|\bra{0}\braket{\Phi_0|w_s}|$ and $\|P_\delta\ket{w_t}\|$.
In what follows, we identify the points of our previous analysis, where we had to use loose inequalities owing to our ignorance of the entire spectrum, and we improve our analysis starting from those points.

First, we note that the second largest eigenvalue of the adjacency matrix is:
\begin{equation}
\cos\phi_1 = \frac{1}{D}(D-1+\cos\frac{2\pi}{L}) = 1-\frac{2}{D}\sin^2\frac{\pi}{L} ~,
\end{equation}
which means that $\phi_1 = \Theta(\frac{1}{L}) = \Theta(N^{-1/D})$ and $g = \frac{2}{D}\sin^2(\frac{\pi}{L}) = \Theta(N^{-2/D})$.

\paragraph{Lower bound on $\alpha_\delta$:} 
We start from Eq.\eqref{eq:mult_eval0_tulsi} to get,
\begin{align}
\label{eq:mult_master_latt}
(\frac{M}{N} ~\text{cosec}^2\frac{\alpha}{2} - \tan^2 \delta\sec^2\frac{\alpha_\delta}{2}) \sum_{\mathbf{y} \in T} x_{\mathbf{y}} &= 2\sum_{\mathbf{k}\ne0} \frac{(\sum_{\mathbf{x} \in T} a_{\mathbf{k}\mathbf{x}}^*) (\sum_{\mathbf{z} \in T} a_{k\mathbf{z}} x_{\mathbf{z}})}{\cos\alpha_\delta-\cos\phi_\mathbf{k}} ~,\\
  &= \frac{2}{N}\sum_{\mathbf{x},\mathbf{z} \in T} (x_{\mathbf{z}}\sum_{\mathbf{k}\ne0} \frac{e^{2\pi i\mathbf{k}\cdot(\mathbf{z}-\mathbf{x})/L}}{\cos\alpha_\delta-\cos\phi_\mathbf{k}}) ~.
\end{align}
We upper bound the right hand side of this result, using a tighter inequality than Eq.\eqref{eq:g_bound},
\begin{equation}\label{eq:cos_bound_latt}
\cos\alpha_\delta-\cos\phi_{\mathbf{k}} > \cos\frac{\phi_{\mathbf{k}}}{2}-\cos\phi_{\mathbf{k}} > \cos^2\frac{\phi_{\mathbf{k}}}{2}-\cos\phi_{\mathbf{k}} = \frac{1}{2} (1-\cos\phi_{\mathbf{k}}) ~,
\end{equation}
and the fact that the Fourier transform of a symmetric positive function is smaller than its value with $\mathbf{k}=0$ in the phase, to obtain
\begin{equation}
(\frac{M}{N} ~\text{cosec}^2\frac{\alpha_\delta}{2} - \tan^2\delta\sec^2\frac{\alpha_\delta}{2}) \sum_{\mathbf{y} \in T} x_{\mathbf{y}} < \frac{4M}{N}\sum_{\mathbf{z} \in T} x_{\mathbf{z}} ~ \sum_{\mathbf{k}\ne0} \frac{1}{1-\cos\phi_\mathbf{k}} ~.
\end{equation}
Eliminating norms of $x_i$, we get
\begin{equation}\label{eq:alpha_bound_latt}
\text{cosec}^2\frac{\alpha_\delta}{2} < 4\sum_{\mathbf{k}\ne0} \frac{1}{1-\cos\phi_\mathbf{k}} + \frac{N}{M}\tan^2\delta\sec^2\frac{\alpha_\delta}{2} ~.
\end{equation}
Since $\sum_{\mathbf{k}\ne0} \frac{1}{1-\cos\phi_{\mathbf{k}}} = \Theta(N)$ for $D>2$, the second term on the right hand side does not exceed the first one for $\tan\delta=O(\sqrt{M})$; the two terms are comparable for $\tan\delta=\Theta(\sqrt{M})$.
Making this choice and using $\sin\frac{\alpha_\delta}{2} < \frac{\alpha_\delta}{2}$, we obtain the lower bound
\begin{equation}
\alpha_\delta = \Omega(\frac{1}{\sqrt{N}}) ~.
\end{equation}
Moreover, the upper bound on $\alpha_\delta$ in Eq.\eqref{eq:alpha_upper} still holds, and so we can attain $\alpha_\delta=\Theta(\frac{1}{\sqrt{N}})$.

For $D>4$, we can improve this result slightly.
Instead of Eq.\eqref{eq:matrix_element} of Lemma \ref{lemma:1}, we have using the Cauchy-Schwarz inequality, 
\begin{equation}\label{eq:matrix_element_latt}
|\frac{M}{N} ~\text{cosec}^2\frac{\alpha_\delta}{2} - \tan^2\delta\sec^2\frac{\alpha_\delta}{2})|~|\sum_{\mathbf{y} \in T} x_{\mathbf{y}}| \le 2\sqrt{N} \|V_{\alpha_\delta} P \ket{\Phi_0}\|_2 ~ \|P\ket{X_\delta}\|_2 ~.
\end{equation}
Now we use our knowledge of the complete spectrum to evaluate $\|V_{\alpha_\delta} P\ket{\Phi_0}\|_2$:
\begin{align}
\|V_{\alpha_\delta} P \ket{\Phi_0}\|^2_2 &= \sum_{\mathbf{k}} |\braket{\Phi_k| V_{\alpha_\delta} P|\Phi_0}|^2 ~,\\
  &= \sum_{\mathbf{k}\ne0} \frac{|\braket{\Phi_k|P|\Phi_0}|^2}{(\cos\alpha_\delta-\cos\phi_{\mathbf{k}})^2} ~,\\
  &= \frac{1}{N} \sum_{\mathbf{k}\ne0} \sum_{\mathbf{y} \in T} \frac{|a_{\mathbf{ky}}|^2}{(\cos\alpha_\delta-\cos\phi_{\mathbf{k}})^2} ~,\\
  &= \frac{2M}{N^2} \sum_{\mathbf{k}>0} \frac{1}{(\cos\alpha_\delta-\cos\phi_{\mathbf{k}})^2} ~.
\label{Valpha_latt_bound}
\end{align}
Then, using Eq.\eqref{eq:cos_bound_latt}, we obtain the bound,
\begin{equation}
\|V_{\alpha_\delta} P \ket{\Phi_0}\|^2_2 < \frac{8M}{N^2} \sum_{\mathbf{k}>0} \frac{1}{(1-\cos\phi_{\mathbf{k}})^2} ~.
\end{equation} 
In Appendix E, we show that $\sum_{\mathbf{k}\ne0} \frac{1}{(1-\cos\phi_{\mathbf{k}})^2} = \Theta(N)$ for $D>4$, which makes
\begin{equation}
\|V_{\alpha_\delta} P \ket{\Phi_0}\|^2_2 = O(\frac{M}{N}) ~.
\end{equation} 
Substituting it in Eq.\eqref{eq:matrix_element_latt}, we have instead of Eq.\eqref{eq:mult_eval_1} of Lemma \ref{lemma:1},
\begin{equation}\label{eq:mult_eval_latt}
0 < \frac{M}{N} ~ \text{cosec}^2\frac{\alpha_\delta}{2} - \tan^2\delta\sec^2\frac{\alpha_\delta}{2} = O(\frac{\sqrt{M}}{R}) ~.
\end{equation}

Proceeding as in Lemma \ref{lemma:1}, the left inequality in Eq.\eqref{eq:mult_eval_latt} gives the same upper bound on $\alpha_\delta$ as in Eq.\eqref{eq:alpha_upper}.
The right inequality, on the other hand, can be rearranged as
\begin{equation}
\frac{1}{\sin\alpha_\delta} < O\left( \frac{1}{\tan\delta}\sqrt{\frac{N}{M}(2-\cos\alpha_\delta)} \right) + \frac{N}{2M}\tan^2\delta\tan\frac{\alpha_\delta}{2}\sec^2\frac{\alpha_\delta}{2} ~.
\end{equation}
With the choice $\tan\delta=\Theta(1)$, which implies $\alpha_\delta=O(\sqrt{\frac{M}{N}})$, both the terms on the right hand side scale as $\Theta(\sqrt{\frac{N}{M}})$, and we arrive at
\begin{equation}
\alpha_\delta = \Theta(\sqrt{\frac{M}{N}}) ~.
\end{equation}
This result is an improvement over that of Lemma \ref{lemma:1} by a factor of $\sqrt{g}$.

We note that the condition $\alpha_\delta<\phi_1/2$, required in Lemmas \ref{lemma:2} and \ref{lemma:3}, is automatically satisfied for $\alpha_\delta=\Theta(\frac{1}{\sqrt{N}})$, since $\phi_1=\Theta(\sqrt{g})=\Theta(N^{-1/D})$ and $D>2$.
For $\alpha_\delta=\Theta(\sqrt{\frac{M}{N}})$ and $D>4$, the condition requires $M=o(N^{(D-2)/D})$, which we shall assume whenever necessary.

\paragraph{Lower bound on $|\braket{\tilde{\Phi}_0|w_s}|^2$:} 
For $\alpha_\delta=\Theta(\frac{1}{\sqrt{N}})$ and $D>2$, we have from Lemma \ref{lemma:2}, 
\begin{equation}
\frac{1}{|\braket{\tilde{\Phi}_0|w_s}|^2} = 1 + O(N^{(2-D)/D}) ~.
\end{equation}
This gives $|\braket{\tilde{\Phi}_0|w_s}|^2 = \Theta(1)$.
The same result also holds for $\alpha_\delta=\Theta(\sqrt{\frac{M}{N}})$ and $D>4$, because $M=o(N^{(D-2)/D})$.

\paragraph{Lower bound on $\|P_\delta\ket{w_t}\|^2$:}
First we consider $\alpha_\delta=\Theta(\frac{1}{\sqrt{N}})$, $\tan\delta=\Theta(\sqrt{M})$ and $D>2$.
Then we get from Eq.\eqref{eq:pwt_bound4},
\begin{equation}
\frac{1}{\|P_\delta\ket{w_t}\|^2} = \Theta(1) ~.
\end{equation}
yielding $\|P_\delta\ket{w_t}\|^2 = \Theta(1)$.
The same result holds for $\alpha_\delta=\Theta(\sqrt{\frac{M}{N}})$, $\tan\delta=\Theta(1)$ and $D>4$ too.
Hence, in either case, after $Q = \lfloor\frac{\pi}{2\alpha_\delta}\rfloor$ iterations, the search algorithm succeeds with probability $\Theta(1)$.

All put together, for $2<D\le4$, the oracle complexity is $\Theta(\sqrt{N})$, which is an improvement over the result of Theorem \ref{theo:3} by a factor of $\sqrt{gM}$, and is optimal as far as the dependence on $N$ is concerned.%
\footnote{Of course, the improvement would be relevant only if $gM$ is smaller than $1$, i.e. $M=O(N^{2/D})$.}
~For $D>4$, the oracle complexity is $\Theta(\sqrt{\frac{N}{M}})$, which is an improvement over the result of Theorem \ref{theo:3} by a factor of $\sqrt{g}$, and is optimal.

Our analysis can be easily extended to the case $D=2$, where $\sum_{\mathbf{k}\ne0} \frac{1}{1-\cos\phi_{\mathbf{k}}} = \Theta(N\log N)$.
Then Eq.\eqref{eq:alpha_bound_latt} leads to $\alpha_\delta=\Theta(\frac{1}{\sqrt{N\log N}})$ with $\tan\delta=\Theta(\sqrt{M\log N})$.
The overlaps $|\braket{\tilde{\Phi}_0|w_s}|^2$ and $\|P_\delta\ket{w_t}\|^2$ remain $\Theta(1)$, and the oracle complexity becomes $\Theta(\sqrt{N\log N})$.
A summary of our results on $D$-dimensional hypercubic lattices is given in Table~\ref{tab:srch_param_latt}.

\smallskip
\begin{table}[th]
\begin{center}
\tcaption{\label{tab:srch_param_latt}Our best results for the parameters of the QGS algorithm, in case of spatial search on $D$-dimensional hypercubic lattices.}
\medskip
\renewcommand{\arraystretch}{1.5}
\begin{tabular}{|c|c|c|c|}
  \hline 
  Dimension & $\tan\delta$             & $\alpha_\delta$                    & $Q_\delta$ \\ 
  \hline 
  $D=2$     & $\Theta(\sqrt{M\log N})$ & $\Theta(\frac{1}{\sqrt{N\log N}})$ & $\Theta(\sqrt{N\log N})$ \\ 
  $2<D\le4$ & $\Theta(\sqrt{M})$       & $\Theta(\frac{1}{\sqrt{N}})$       & $\Theta(\sqrt{N})$ \\ 
  $D>4$     & $\Theta(1)$              & $\Theta(\sqrt{\frac{M}{N}})$       & $\Theta(\sqrt{\frac{N}{M}})$ \\ 
  \hline 
\end{tabular}   
\end{center}
\end{table}  

\smallskip
It has been observed that the presence of multiple targets in certain special configurations can cause spatial search to fail \cite{nahi2017walk}.
The problem caused by such exceptional configurations depends on the nature of the oracle.
It is absent in our algorithm, because the oracle chosen by us is different from the one considered in Ref.~\cite{nahi2017walk}.

\section{Quantum Bounds on Classical Hitting Time}
\label{sec:Quantum_bounds}

We next use our results to derive some bounds on the classical hitting times of random walks on regular graphs.
The \textit{hitting time} or \textit{access time} between two vertices $u$ and $v$ is defined as the expected number of steps before a random walker starting from $u$ reaches $v$.
The {\it average hitting time}, $h_T$, is the expected number of steps needed for a random walker starting with a uniform distribution to reach the marked set $T$.
$h_T$ decreases with increasing graph connectivity, and reaches its minimum for the maximally connected complete graph.
For $M$ targets on the complete graph, $h_T=\frac{(N-M)(N-1)}{MN}$.
For a single target on the $D$-dimensional hypercubic lattice, $h_T$ is $\Theta(N)$ for $D>2$ and $\Theta(N\log N)$ for $D=2$ \cite{szegedy2004quantum}.
It is possible to construct graphs for which the average hitting time is $\Theta(N^3)$ \cite{lovasz1993random}, but such graphs are generally poorly connected and have small values of $g$.

General good upper bounds on $h_T$ are hard to deduce, and it is worthwhile to see how information about the graph connectivity can improve the situation.
The following result by Szegedy \cite{szegedy2004quantum,szegedy2004spectra} on the average hitting time of a graph incorporates the connectivity aspect of the graph as well.

\smallskip
\begin{lemma}
\label{lemma_szegedy}
Let $\alpha$ be the smallest eigenphase of the leaking walk matrix and let $\vec{\alpha}$ be the corresponding normalized eigenvector.
Then $h_T = O(\frac{1}{\alpha^2})$ and $h_T = \Omega(\frac{\|\alpha\|^2_1}{N}\frac{1}{\alpha^2})$.
\end{lemma}

The adjacency matrix has the uniform eigenvector with eigenvalue $1$, and $\|\alpha\|^2_1$ describes how uniform the corresponding vector of the leaking walk matrix is.
Intuitively, $\|\alpha\|^2_1$ measures the irregularity of the graph after the target states have been removed.
If the removal of the target vertices does not disturb the regularity of the adjacency matrix too much, then $\|\alpha\|^2_1$ would scale as $\Theta(N-M)$.
This property is mentioned by Szegedy \cite{szegedy2004quantum}, but not proven.
We can use Lemma \ref{lemma:2} to show that this is indeed the case for regular graphs with large enough $g$, which is tantamount to sufficiently large connectivity,

\smallskip
\begin{lemma}
\label{lemma:uniform}
Consider a regular graph of size $N$, spectral gap $g$, and a marked set of size $M$.
Then, provided that $\alpha<\phi_1/2$,
\begin{equation}
\|\alpha\|_1 > \sqrt{N}(1-O(\frac{\alpha^2}{g})) ~.
\end{equation}
\end{lemma}

\noindent
This result follows from Lemma \ref{lemma:2} by setting $\delta = 0$ and $\|\alpha\|_1=\sqrt{N}|\langle w_s|\Phi_0\rangle|$, the proof of which is given in Appendix F.
It strengthens Szegedy's lemma to show that $h_T$ is exactly characterized by $\frac{1}{\alpha^2}$, for regular graphs with large enough connectivity.

\smallskip
\begin{lemma}
\label{lemma:hitting}
Consider a regular graph of size $N$, spectral gap $g$, and a marked set of size $M$.
Then, provided that $g>2\sin^2\alpha$, the average hitting time associated with the marked set is, $h_T = \Theta(\frac{1}{\alpha^2})$, where $\cos\alpha$ is the principal eigenvalue of the leaking walk matrix. 
\end{lemma}

\smallskip\noindent
Note that the condition $g>2\sin^2\alpha$ is equivalent to $\alpha<\phi_1/2$, and makes $O(\alpha^2/g)=O(1)$.
Also, $N$ and $M$ do not explicitly appear in the result.
There is nothing inherently quantum mechanical about this result, but the application of quantum ideas makes it derivation simple.
Several proofs of this nature are reviewed by Drucker and de Wolf \cite{dewolf2009proof}.

\section{Summary}

In this work, we have extended and put on a rigorous footing several earlier observations regarding spatial search using quantum walk on regular graphs.
In Section \ref{sec:eigenWU}, we have explicitly related the eigenvalues and eigenvectors of the adjacency matrix $A$ to those of the flip-flop walk operator $W$.
The classical diffusion equation, $f(t+\Delta t) = A f(t)$, has the spectrum described by the non-relativistic dispersion relation $\omega(\phi)=1-\cos\phi$, while the quantum evolution equation $f(t+\Delta t) = W f(t)$, has the spectrum described by the relativistic dispersion relation $\omega(\phi)=\pm\phi$.
The flip-flop walk thus happens to be a simple way to quadratically speed up information propagation around the graph, by converting a classical non-relativistic evolution to a quantum relativistic one.
Moreover, our explicit counting of the degrees of freedom in Appendix C shows that the relativistic dynamics generated by this process is that of the complex Klein-Gordon field.
In particular, the modes with eigenvalues $e^{\pm i\phi}$ can be interpreted as particle-antiparticle mode pairs.

In Section \ref{sec:QGSanalysis}, we have analysed the Quantum Graph Search algorithm with multiple targets, and showed that for regular graphs with spectral gap $g$, the algorithm yields one of the target states with $\Theta(1)$ probability using $Q=O(\sqrt{\frac{N}{gM}})$ oracle calls.
In this process, we have extended the abstract search framework introduced in Ref.~\cite{ambainis2005coins} to the case of multiple target items, and used Tulsi's controlled spatial search technique \cite{tulsi2008faster} to speed up the algorithm.
Our analysis assumes neither any specific structure for the graph (except the spectral gap) nor any specific locations for the target items.
As a result, there is room for improvement in the bounds we have proved, when more information is available about the graph structure.

Our proofs of the technical lemmas show that the actual computational complexity of the QGS depends on the complete spectrum of the adjacency matrix as well as the leaking walk matrix.
Consequently, established classical results on the nature of these matrices, for a variety of graphs, can be directly used to improve the performance of the QGS.
We have worked out two special cases explicitly, where the eigenspectra of the graphs are known: the QGS for the complete graph in Section \ref{sec:QGScomplete}, and the QGS for the hypercubic lattice in Section \ref{sec:QGSlattice}.
For the complete graph, the oracle complexity of the QGS is inferior to that for the optimal Grover search by a factor of two.
For hypercubic lattices of dimension $D>2$, the oracle complexity of the QGS reaches the optimal scaling behaviour as a function of the size of the graph; it attains optimal scaling behaviour as a function of the number of targets too, for $D>4$.
Numerical simulations on hypercubic lattices have shown that the oracle complexity of the QGS depends not only on the number of targets but also on how they are distributed around the lattice \cite{patel2010search2,patel2010search}.
Our analysis is not accurate enough to capture these features, and how to modify it to capture them would be a desirable extension of our work.

Our work also gives improved bounds on classical average hitting time of random walks on regular graphs, which take into account the graph connectivity.
We show, by building on the work of Szegedy, that the average hitting time is characterized by the principal eigenvalue of the leaking walk matrix for well-connected graphs.

\appendix{: Eigenvalues of $W$}

\noindent
In the appendices, we prove some auxiliary results that are necessary for the proofs in the main part of this article.
We first look at the multiplicities of the real eigenvalues of $W$.

\smallskip\noindent
{\bf Lower bound on the multiplicity of $1$:}
$W$ is a real unitary matrix, so its real eigenvalues can only be $1$ or $-1$, and its complex eigenvalues must come in conjugate pairs.
The number of complex eigenvalues are fixed by Theorem 1 to be $2N-2$.
Now let us count the $1$-eigenvectors of $W$.
One way a vector can be a $1$-eigenvector of $W$ is if it is a $-1$-eigenvector of both $C$ and $S$.
Both $C$ and $S$ are reflection operators, so we know their eigenspaces directly from their definitions.
For a vector $\ket{\phi}$ to be a $-1$-eigenvector of $C$, it has to be orthogonal to $\ket{H,v}$ for all $v \in V$.
This imposes a set of $N$ conditions on the components of $\ket{\phi}$.
Similarly, for $\ket{\phi}$ to be a $-1$-eigenvector of $S$, it has to be orthogonal to $\ket{e_+}$ for all $e \in E$.
This imposes another $dN/2$ conditions on the components of $\ket{\phi}$.
These two sets of conditions may be linearly dependent, so together they impose at most $(dN/2)+N-1$ conditions on the components of $\ket{\phi}$.
(We have subtracted $1$ from the number of conditions, because we know that $\ket{\Phi_0}$ is a $1$-eigenvector for both $C$ and $S$.)
Consequently, at least $Nd-((Nd/2)+N-1)$ components of $\ket{\phi}$ can be freely chosen, giving at least $(Nd/2)-N+1$ linearly independent $1$-eigenvectors for $W$.
We add $\ket{\Phi_0}$ to this list, as it is a $1$-eigenvector of both $C$ and $S$.
So there are at least $(Nd/2)-N+2$ linearly independent $1$-eigenvectors for $W$.

\smallskip\noindent
{\bf Lower bound on the multiplicity of $-1$:}
In the same vein, we find a lower bound for the number of $-1$-eigenvectors of $W$, by counting the number of vectors which are $-1$-eigenvectors of $C$ and $1$-eigenvectors of $S$.
As before, the requirement that $\ket{\phi}$ is a $-1$ eigenvector of $C$ imposes $N$ conditions on the components of $\ket{\phi}$.
Also, to be a $1$-eigenvector of $S$, $\ket{\phi}$ must be orthogonal to all $\ket{e^-}$ states.
This imposes another $dN/2$ conditions on the components of $\ket{\phi}$.
Possible linear dependence between these conditions means that their number is at most $(Nd/2)+N$.
So at least $(Nd/2)-N$ components of $\ket{\phi}$ can be freely chosen, implying that there are at least $(Nd/2)-N$ linearly independent $-1$-eigenvalues for $W$.

We thus have lower bounds for the multiplicities of $1$ and $-1$ eigenvalues of $W$.
But we know that there are exactly $Nd-2N+2$ real eigenvalues of $W$, from Corollary \ref{cor:1}, when $G$ is not bipartite.
As a result, these lower bounds give the actual multiplicities of $1$ and $-1$ eigenvalues of $W$.

\smallskip\noindent
{\bf Bipartite graphs:}
Similar results follow for bipartite graphs.
Since the adjacency matrix of a bipartite graph has a $-1$ eigenvalue, $W$ has $2(N-2)$ complex eigenvalues as per Corollary \ref{cor:1}.
That makes the lower bound on the number of real eigenvalues loose for bipartite graphs, with two real eigenvalues unaccounted for.
We can make the lower bound on the number of $-1$ eigenvalues tight, by noting that $\ket{\Phi_b}$ is a $-1$-eigenvector of $W$ for bipartite graphs.
Then, $\ket{\Phi_b}$ plays the same role in obtaining the lower bound on the multiplicity of $-1$ eigenvalue, as $\ket{\Phi_0}$ did in obtaining the lower bound on the multiplicity of $1$ eigenvalue.
$\ket{\Phi_b}$ is a $1$ eigenvector of $C$ and a $-1$ eigenvector of $S$.
So the number of $-1$-eigenvectors of $W$ is increased by two, because $\ket{\Phi_b}$ is added to this list and the number of independent conditions that we counted before is decreased by one.
Hence, there are at least $(Nd/2)-N+2$ linearly independent $-1$-eigenvectors for $W$.
Overall, the multiplicity of $1$ eigenvalue remains unchanged at $(Nd/2)-N+2$, and the two extra real eigenvalues are both $-1$.

\appendix{: Tulsi's controlled spatial search algorithm}

\noindent
Given a quantum search algorithm, defined in terms of the spatial search operator $U=WO$, Tulsi's method improves its convergence by controlling the operations using an additional ancilla bit \cite{tulsi2008faster}.
This controlled spatial search algorithm uses a tunable parameter $\delta$, and can be implemented using the quantum logic circuit shown in Fig.~\ref{fig:tulsi_circ}.

The starting state for the algorithm is $\ket{\psi_s} = \ket{\Phi_0}\ket{0}$.
The single qubit operators are:
\begin{equation}
X_\delta = \begin{pmatrix}
~\cos\delta & \sin\delta \\
-\sin\delta & \cos\delta
\end{pmatrix} ,~~
Z = \begin{pmatrix} 1 & 0 \\ 0 & -1 \end{pmatrix} .
\end{equation}
After $Q_\delta$ iterations of the search operator, enclosed in the dashed box, the algorithm reaches a final state that has $\Theta(1)$ overlap with the target state $\ket{\psi_i}\ket{\delta}$, where $\ket{\delta}=X_\delta^\dagger\ket{0}$.

\begin{figure}[th]
{
\begin{center}

\begin{picture}(180,90)
\put(20,60){\line(1,0){18}}
\put(54,60){\line(1,0){10}}
\put(68,60){\line(1,0){10}}
\put(94,60){\line(1,0){10}}
\put(108,60){\line(1,0){10}}
\put(134,60){\line(1,0){16}}
\put(20,20){\line(1,0){40}}
\put(72,20){\line(1,0){28}}
\put(112,20){\line(1,0){38}}

\put(66,60){\circle{4}}
\put(106,60){\circle{4}}
\put(66,26){\line(0,1){32}}
\put(106,26){\line(0,1){32}}

\put(38,52){\framebox(16,16){$X_\delta$}}
\put(78,52){\framebox(16,16){$X_\delta^\dagger$}}
\put(118,52){\framebox(16,16){$Z$}}
\put(60,14){\framebox(12,12){$O$}}
\put(100,14){\framebox(12,12){$W$}}

\put(0,18){$|\Phi_0\rangle$}
\put(4,57){$|0\rangle$}
\put(154,57){$|\delta\rangle$}
\put(154,18){$|\psi_i\rangle$}

\put(30,10){\dashbox{5}(110,65){}}
\put(90,80){\makebox(0,0)[b]{Iterate $Q_\delta$ times}}
\end{picture}
\end{center}
}
\fcaption{\label{fig:tulsi_circ}Quantum logic circuit for Tulsi's controlled spatial search algorithm.}
\end{figure}

\appendix{: Invariant subspaces of $U$ and $U_\delta$}

\noindent
We define the subspace $\mathcal{H} \subset \mathbb{C}^{Nd}$, as the $2N-1$ dimensional subspace spanned by all the eigenvectors corresponding to non-real eigenvalues of $W$ and $\ket{\Phi_0}$.
Notice that all vectors of the form $\ket{H,v}$, which are $1$-eigenvectors of $C$ as well as eigenvectors of $O$, lie entirely inside $\mathcal{H}$.
This is because all real eigenvectors of $W$, except $\ket{\Phi_0}$, are also $-1$-eigenvectors of $C$; so by definition they are perpendicular to all vectors of the form $\ket{H,v}$.
We use this fact to prove that $\mathcal{H}$ is invariant under the action of $U$ for non-bipartite graphs.

Similarly, with the ancilla included, we define the subspace $\tilde{\mathcal{H}} \subset \mathbb{C}^{2Nd}$, as the $2N+M-1$ dimensional subspace spanned by all the eigenvectors corresponding to non-real eigenvalues of $\tilde{W}$, $\ket{\Phi_0}\ket{0}$ and $\ket{\psi_{i \in T}}\ket{1}$. 
In this case, all vectors of the form $\ket{H,v}\ket{0}$, which are $1$-eigenvectors of $\overline{\rm ctrl}$-$CZ$, lie entirely inside $\mathcal{H}$.
This is because all real eigenvectors of $\tilde{W}$, except $\ket{\Phi_0}\ket{0}$, are also $-1$-eigenvectors of $\overline{\rm ctrl}$-$CZ$; so by definition they are perpendicular to all vectors of the form $\ket{H,v}\ket{0}$.
This fact leads to the proof that $\tilde{\mathcal{H}}$ is invariant under the action of $U_\delta$ for non-bipartite graphs.

\smallskip
\begin{lemma} 
$U(\mathcal{H}) = \mathcal{H}$ and $U_\delta(\tilde{\mathcal{H}}) = \tilde{\mathcal{H}}$.
\end{lemma}

\noindent{\bf Proof:}
Let $\ket{\phi}$ be any vector in $\mathcal{H}$.
Now, $U\ket{\phi} = W\ket{\phi} - 2\sum_{i \in T} W\ket{\psi_i} \braket{\psi_i|\phi}$.
$\mathcal{H}$ is by construction invariant under $W$, so $W\ket{\phi}\in\mathcal{H}$.
For any $i \in T$, $\ket{\psi_i}$ is a vector of the form $\ket{H,v}$.
That means both $\ket{\psi_i}$ and $W\ket{\psi_i}$ belong to $\mathcal{H}$.
Since $U_\ket{\phi}$ is a linear combination of the vectors $W\ket{\phi}$ and $W\ket{\psi_i}$, it also belongs to $\mathcal{H}$.

Similarly, for any vector $\ket{\phi}$ in $\tilde{\mathcal{H}}$, $U_\delta\ket{\phi} = \tilde{W}\ket{\phi} - 2\sum_{i \in T} \tilde{W}\ket{\psi_{i,\delta}} \braket{\psi_{i,\delta}|\phi}$.
Also, invariance of $\tilde{\mathcal{H}}$ under $\tilde{W}$ makes $\tilde{W}\ket{\phi}\in\tilde{\mathcal{H}}$.
Now, for any $i \in T$ and all $\delta$, $\ket{\psi_{i,\delta}}$ is a linear combination of vectors of the form $\ket{H,v}\ket{0}$ and $\ket{\psi_i}\ket{1}$.
That means both $\ket{\psi_{i,\delta}}$ and $\tilde{W}\ket{\psi_{i,\delta}}$ belong to $\tilde{\mathcal{H}}$.
Finally, $U_\delta\ket{\phi}$ is a linear combination of the vectors $\tilde{W}\ket{\phi}$ and $\tilde{W}\ket{\psi_{i, \delta}}$, and so it belongs to $\tilde{\mathcal{H}}$.
\hfill $\square\,$ 

\smallskip
To extend this result to bipartite graphs, we have to define $\mathcal{H}$ as the $2N-2$ dimensional subspace spanned by $\ket{\Phi_0}$, $\ket{\Phi_b}$ and all the eigenvectors corresponding to non-real eigenvalues of $W$.
The same lemma then holds, because all real eigenvectors of $W$, except for $\ket{\Phi_0}$ and $\ket{\Phi_b}$, are $-1$ eigenvectors of $C$, and hence all vectors of the form $\ket{H,v}$ lie in $\mathcal{H}$.
Similarly, for bipartite graphs with the ancilla included, $\tilde{\mathcal{H}}$ is the $2N+M-2$ dimensional subspace spanned by $\ket{\Phi_0}\ket{0}$, $\ket{\Phi_b}\ket{0}$, $\ket{\psi_{i \in T}}\ket{1}$ and all the eigenvectors corresponding to non-real eigenvalues of $\tilde{W}$.
Then the lemma holds, because all real eigenvectors of $\tilde{W}$, except for $\ket{\Phi_0}\ket{0}$ and $\ket{\Phi_b}\ket{0}$ are $-1$ eigenvectors of $\overline{\rm ctrl}$-$CZ$, and hence all vectors of the form $\ket{H,v}\ket{0}$ lie in $\tilde{\mathcal{H}}$.

As a byproduct of this analysis, the dimensionality of $\mathcal{H}$ lets us infer that the degrees of freedom for $\mathcal{H}$ actually represent the propagating modes of the complex Klein-Gordon field on $N$ vertices.
This feature has been noted in case of hypercubic lattices in Ref.~\cite{patel2012search}.
The modes associated with global symmetries are real---the eigenvectors $\ket{\Phi_0}$ and $\ket{\Phi_b}$ are associated with translation and reflection symmetries respectively---and that accounts for the dimensionality of $\mathcal{H}$ being less than $2N$.%
\footnote{Unitary transformations (i.e. phase changes) corresponding to symmetry directions do not change the physical state of a quantum system.}

\appendix{: Quantum search with $\delta=0$}

\noindent
Quantum search without any ancilla control, using $U = WO$, produces subpar results.
To demonstrate it, we first bound $\alpha$, using classical graph theory.
(Our proof of Lemma \ref{lemma:1} cannot be applied to the case $\delta=0$.)

\smallskip
\begin{lemma}
\label{lemma:6}
Let $e^{i\alpha}$ be the eigenvalue of $U$ closest to $1$.
Then in terms of the spectral gap of the adjacency matrix $g$, the number of graph vertices $N$, and the number of target vertices $M$, we have $\frac{\pi}{\sqrt{2}}\sqrt{\frac{M}{N}} > \alpha > \sqrt{\frac{gM}{N}}$.
\end{lemma}

\paragraph{Proof of upper bound:}
For any real symmetric $N \times N$ matrix $\mathbf{M}$, the largest eigenvalue $\lambda_{max}$ can be characterized as,
\begin{equation}
\lambda_{max} = \underset{\vec{x}\in\mathbb{R}^N}{\max} ~ \frac{\vec{x}\cdot\mathbf{M}\vec{x}}{\vec{x}\cdot\vec{x}} ~.
\end{equation}  
From Theorem \ref{theo_2}, we know that $\cos\alpha$ is the largest eigenvalue of $\tilde{A}_T$.
So
\begin{equation}
\cos\alpha = \underset{\vec{x}\in\mathbb{R}^{N-M}}{\max} ~ \frac{\vec{x}\cdot\tilde{A}_T\vec{x}}{\vec{x}\cdot\vec{x}}.
\end{equation}

Now, consider the particular $\vec{y}=(1,1,\ldots,1)\in\mathbb{R}^{N-M}$.
Then $\vec{y}\cdot\vec{y} = N-M$.
Also, $\vec{y}\cdot\tilde{A}_T\vec{y}$ is the sum of all the entries in the leaking walk matrix, which is equal to the number of edges in $V-T$ multiplied by $2/d$.
Putting these together,
\begin{equation}
\cos\alpha \ge \frac{2}{d}~\frac{|E(V-T,V-T)|}{N-M} ~.
\end{equation} 
The original graph has $Nd/2$ edges.
Removing the target vertices removes at most $Md$ edges (exactly $Md$ edges are removed if the target vertices do not share any edges).
So $|E(V-T,V-T)| \ge (Nd/2)-Md$, and
\begin{equation}
\cos\alpha \ge \frac{N-2M}{N-M} = 1-\frac{M}{N-M} ~.			
\end{equation}
Using the inequality $1-\frac{2\alpha^2}{\pi^2} \ge \cos\alpha$, we obtain
\begin{equation}
\alpha \le \frac{\pi}{\sqrt{2}}~\sqrt{\frac{M}{N-M}}
       = O(\sqrt{\frac{M}{N}}) ~.
\end{equation}

\paragraph{Proof of lower bound:}
We follow Szegedy \cite{szegedy2004spectra}, and define $\vec{\alpha}^\prime\in \mathbb{R}^N$ as the normalized vector $\vec{\alpha}$ augmented by zeroes at the target vertex locations.
According to the Perron-Frobenius theorem, components of $\vec{\alpha}$ are strictly positive, so
\begin{equation}
\|A\vec{\alpha}'\|^2 > \|\tilde{A}_T\vec{\alpha}\|^2 = \cos^2\alpha ~.
\label{eq:Aalpha_bound1}
\end{equation}

We can also decompose $\vec{\alpha}^\prime$ in the complete orthonormal eigenbasis of $A$, as
\begin{equation}
\vec{\alpha}^\prime = \beta_0\vec{a}_0 + \sum_{k>0}\beta_k\vec{a}_k ~,~~ \beta_0^2 + \sum_{k>0}\beta_k^2 = 1 ~.
\end{equation}
Then
\begin{equation}
\beta_0 = \vec{\alpha}^\prime \cdot \vec{a}_0 \le \|\vec{\alpha}\| \|\vec{y}\|/\sqrt{N} = \sqrt{1-\frac{M}{N}} ~~\text{and}~~
\sum_{k>0}\beta_k^2 = 1-\beta_0^2 \ge \frac{M}{N} ~.
\end{equation}
We therefore obtain
\begin{align}
\|A\vec{\alpha}^\prime\|^2 &= \|\beta_0\vec{a}_0 + \sum_{k>0}\cos(\phi_k)~\beta_k\vec{a}_k\|^2 ~,\\
\label{eq:cosphi_bound}
  &= \beta_0^2 + \sum_{k>0}\cos^2(\phi_k)~\beta_k^2 ~,\\
\label{eq:1-g_bound}
  &\le \beta_0^2 + (1-g)^2\sum_{k>0}\beta_k^2 ~,\\
  &\le \beta_0^2 + (1-g)\sum_{k>0}\beta_k^2 ~,\\
  &= 1 - g\sum_{k>0}\beta_k^2 ~,\\
  &\le 1 - g\frac{M}{N} ~.
\label{eq:Aalpha_bound2}
\end{align}

Combining Eqs.\eqref{eq:Aalpha_bound1} and \eqref{eq:Aalpha_bound2}, we get
\begin{equation}
\cos^2{\alpha} < 1-\frac{gM}{N} ~~\text{or}~~ \sin^2{\alpha} > \frac{gM}{N} ~.
\end{equation}
Using the inequality $\sin(\alpha) < \alpha$, we have
\begin{equation}
\alpha > \sqrt{\frac{gM}{N}} ~,
\label{eq:alpha_lower_bound}
\end{equation} 
which is the bound we want.

This derivation by Szegedy \cite{szegedy2004spectra} is not rigorous for certain exceptional cases:\\
(i) $\cos^2(\phi_k) \le (1-g)^2$ cannot be used in Eq.\eqref{eq:cosphi_bound} when $\phi_k>\pi-\phi_1$, e.g $\phi_k=\pi$ occurs for bipartite graphs.\\
(ii) $(1-g)^2 \le (1-g)$ cannot be used in Eq.\eqref{eq:1-g_bound} when $g>1$, as is the case for the complete graph.\\
We can bypass these exceptions by working with $(\mathbb{I}+A)/2$ and $(\mathbb{I}+\tilde{A}_T)/2$, which have the same eigenvectors while the eigenvalues are shifted to the range $[0,1]$.
Then $\cos\alpha$ is replaced by $(1+\cos\alpha)/2=\cos^2(\alpha/2)$, $g$ is replaced by $g/2$, and we have the bound
\begin{equation}
\label{eq:cosalpha_bound}
\cos^2(\alpha/2) < \sqrt{1-\frac{gM}{2N}} < 1-\frac{gM}{4N} ~,
\end{equation}
which gives the same result as Eq.\eqref{eq:alpha_lower_bound}.
\hfill $\square\,$ 

\smallskip
Comparing Lemma \ref{lemma:6} with Lemma \ref{lemma:1}, we see that the lower bound for $\delta=0$ has the same scaling as that for the controlled algorithm, while the upper bound is worse by a factor of $\sqrt{g}$.
Our proofs of Lemma \ref{lemma:2} and Lemma \ref{lemma:3} apply to the case $\delta=0$.
So next we evaluate those overlaps with the changed bounds on $\alpha$.
The result of Lemma \ref{lemma:2} changes to,
\begin{equation}
\frac{1}{|\braket{\Phi_0|w_s}|^2} < 1 + \frac{\alpha^2}{g} = 1 + O(\frac{M}{Ng}) ~.
\end{equation} 
On the other hand, Eq.\eqref{eq:pwt_bound4} becomes,
\begin{equation}
\frac{1}{\|P\ket{w_t}\|^2} = \cot^2\frac{\alpha}{2}(\frac{2M}{N}) = O(\frac{1}{g}) ~,
\end{equation}
yielding $\|P\ket{w_t}\|^2 = \Omega(g)$.

These results for quantum search without ancilla control are inferior to the controlled search versions on three counts.
First, the overlap $|\braket{\Phi_0|w_s}|$ is not close to $1$ for small values of $g$, because the upper bound on $\alpha$ is independent of $g$ when $\delta=0$.
Second, the condition $\alpha<\phi_1/2$ is not always satisfied when $\delta=0$, which is again related to the loose upper bound on $\alpha$.
Third, and the most concerning issue, is that the quantum search succeeds with a probability of only $\Omega(g)$, as per the logic of Theorem \ref{theo:3}.
That requires addition of amplitude amplification to the algorithm to boost the final success probability to $\Theta(1)$ \cite{brassard2002quantum}.
Consequently, the oracle complexity of the algorithm is $O(\frac{1}{\sqrt{g}\alpha})=O(\frac{1}{g}\sqrt{\frac{N}{M}})$.

Thus the quantum search without ancilla control is slower by a factor of $\sqrt{g}$ compared to the controlled search, and would work well only for graphs with a large spectral gap.
We note, however, that a random $d$-regular graph is almost always an expander graph, and so this simpler algorithm would work well on a large fraction of graphs.

\appendix{: Bounds on $\sum_{\mathbf{k}\ne0} 1/(1-\cos\phi_{\mathbf{k}})^2$}

\noindent
For a $D$-dimensional lattice, it is well-known that
\begin{equation}
\sum_{\mathbf{k}\ne0} \frac{1}{1-\cos\phi_{\mathbf{k}}} = \begin{cases}
  \Theta(N\log N) ~, &\text{if} ~ D=2 ~,\\
  \Theta(N) ~, &\text{if} ~ D>2 ~.
  \end{cases}
\end{equation}  
Using a similar counting of contributions, we obtain bounds for the sum $\sum_{\mathbf{k}\ne0} 1/(1-\cos\phi_{\mathbf{k}})^2$, which together with the inequality Eq.\eqref{eq:cos_bound_latt}, provides an upper bound for the quadratic norm of the operator $V_\alpha$ defined in Eq.\eqref{eq:V_alpha}.
We observe that this sum is related to the integral $\int d^Dk/k^4$, which is infrared divergent for $D\le4$.
Removal of the $\mathbf{k}=0$ contribution regulates the divergence, and we can obtain bounds for the remaining part.

For $\theta \in [-\pi,\pi]$ the following inequalities hold:
\begin{equation}
\frac{\theta^2}{2} \ge 1-\cos\theta \ge \frac{2\theta^2}{\pi^2} ~.
\end{equation}
Using these inequalities in $D$ dimensions, we find
\begin{equation}
1-\cos\phi_{\mathbf{k}} = \frac1D\sum_i (1 - \cos(\frac{2\pi k_i}{L})) = \frac{c}{DL^2} \|\mathbf{k}\|^2 ~,
\end{equation}
with $c\in[8,2\pi^2]$.
Therefore, \begin{equation} \label{eq:sum_bound}
\sum_{\mathbf{k}\ne0} \frac{1}{(1-\cos\phi_{\mathbf{k}})^2} = \frac{D^2L^4}{c^2} \sum_{\mathbf{k}\ne0} \frac{1}{\|\mathbf{k}\|^4} ~.
\end{equation}

The sum $\sum_{\mathbf{k}\ne0} 1/\|\mathbf{k}\|^4$ goes over the points of a $D$-dimensional hypercubic lattice, with side length $L=N^{1/D}$.
We can choose $k_i \in \{-\lfloor L/2 \rfloor,\ldots,0,\ldots,\lfloor L/2 \rfloor\}$, with a weight $1/2$ for the end-points when $L$ is even (i.e. the graph is bipartite).
This lattice can be divided in to concentric hypercubic shells, with the center at the origin, and inner side length $l\in\{1,2,\ldots,\lfloor L/2 \rfloor\}$.
The $l^\text{th}$ shell has $(2l+1)^D - (2l-1)^D = \Theta(2D(2l)^{D-1})$ points in it, and for every point,
\begin{equation}
l \le \|\mathbf{k}\| \le \sqrt{D}l ~.
\end{equation} 
With these considerations,
\begin{equation}
\sum_{\mathbf{k}\ne0} \frac{1}{\|\mathbf{k}\|^4} = \Theta(l^{D-1} \sum_{l=1}^{\lfloor L/2 \rfloor} \frac{1}{l^4}) = \Theta(\sum_{l=1}^{\lfloor L/2 \rfloor} l^{D-5}) ~,
\end{equation}
where the proportionality constant is between $2^D/D$ and $D2^D$.

The asymptotic behaviour of the one-dimensional sum, $\sum_{l=1}^{\lfloor L/2 \rfloor} l^{D-5}$ depends on the value of $D$, as follows:\\
$D<4$: The sum $\sum_{l=1}^{\infty} l^{D-5}$ is convergent, and provides a constant upper bound to $\sum_{l=1}^{\lfloor L/2 \rfloor} l^{D-5}$.
Since there is a trivial lower bound of $1$, we have $\sum_{l=1}^{\lfloor L/2 \rfloor} l^{D-5} = \Theta(1)$.\\
$D=4$: The sum $\sum_{l=1}^{\lfloor L/2 \rfloor} l^{-1}$ is the $\lfloor L/2 \rfloor^\text{th}$ harmonic number, which is known to be $\Theta(\log(L))=\Theta(\log N)$.\\
$D>4$: By comparing the sum with the integral of $l^{D-5}$, we have the bounds,
\begin{equation}
1 + \int_1^{\lfloor L/2 \rfloor} dl~l^{D-5} \leq \sum_{l=1}^{\lfloor L/2 \rfloor} l^{D-5} \leq \int_1^{\lfloor L/2 \rfloor+1} dl~l^{D-5} ~.
\end{equation}
Evaluating these, we get $\sum_{l=1}^{\lfloor L/2 \rfloor} l^{D-5} = \Theta(L^{D-4}) = \Theta(N^{1-\frac{4}{D}})$.

Inserting these results in Eq.\eqref{eq:sum_bound}, we obtain
\begin{equation}
\sum_{\mathbf{k}\ne0} \frac{1}{(1-\cos\phi_{\mathbf{k}})^2} = \begin{cases}
  \Theta(N^{4/D}) ~, &\text{if} ~ D<4 ~,\\
  \Theta(N\log N) ~, &\text{if} ~ D=4 ~,\\
  \Theta(N) ~, &\text{if} ~ D>4 ~.
  \end{cases}
\end{equation}  
The scaling behaviour is dimension dependent, because for $D<4$ the dominant contribution is from points near the origin, while for $D>4$ the contribution from points far from the origin dominates.

\appendix{: Linear Norm of $\vec{\alpha}$}

\noindent
With $\lambda\rightarrow\alpha$, Eqs. \eqref{eq:edge_overlap}-\eqref{no_overlap-} give the components of $\ket{\alpha}$ in the edge basis.
Using them, we evaluate the norm of the real vector $\vec{\alpha}$ when $\ket{\alpha}$ is a unit vector.
Since $|\alpha\rangle$ and $|-\alpha\rangle$ form an orthonormal complex conjugate pair, the components of the unit vector $\ket{w_s} = \frac{\ket{\alpha} + \ket{-\alpha}}{\sqrt{2}}$ in the edge basis are,

\begin{enumerate}
\item For edges in $V-T$,
\begin{equation}
\braket{w_s | e^\pm} = \sqrt{\frac{1}{d}} (\vec{\alpha}_u \pm \vec{\alpha}_v)
\end{equation}

\item For edges between $V-T$ and $T$,
\begin{equation}
\braket{w_s | e^\pm} =  \sqrt{\frac{1}{d}} (\vec{\alpha}_u)
\end{equation}

\item For edges in $T$,
\begin{equation}
\braket{w_s | e^\pm} =  0
\end{equation}
\end{enumerate}

\noindent
Now, using the fact that the edge basis is a complete basis,
\begin{align}
1 &= ~ \sum_{e \in E,~ e=(u,v)} (\braket{w_s | e^+}^2 + \braket{w_s | e^-}^2),\\
  &= ~ \frac{2}{d} \sum_{e \in E(V-T,V-T),~ e=(u,v)} (\vec{\alpha}^2_u + \vec{\alpha}^2_v) ~ + ~ \frac{2}{d} \sum_{e \in E(V-T,T),~ e=(u,v)} (\vec{\alpha}^2_u), \\
 & = ~ 2 \sum_{u \in V-T} \vec{\alpha}^2_u, 
\end{align}
i.e. the quadratic norm of $\vec{\alpha}$ is $\frac{1}{\sqrt{2}}$.

To obtain the linear norm of $\vec{\alpha}$, we look at $\braket{w_s|\psi_i}$.
For all $i \in T$, according to Corollary \ref{cor:2}, $\braket{\alpha|\psi_i}$ are imaginary and so $\braket{w_s|\psi_i}$ vanish.
Furthermore, for all $i \in V-T$, we have chosen all $\vec{\alpha}_i$ to be real and positive, as mentioned in the proof of Corollary \ref{cor:2}.
Thus we can write the overlap of $\ket{w_s}$ with the uniform superposition state as,
\begin{equation}
|\braket{w_s|\Phi_0}| = \frac{1}{\sqrt{N}}|\sum_{i \in V} \braket{w_s|\psi_i}| = \sqrt{\frac{2}{N}} \sum_{i \in V-T} |\vec{\alpha}_i|.
\end{equation}
Hence, the linear norm $\|\vec{\alpha}\|_1 = \sqrt{N}|\braket{w_s|\Phi_0}$, in the convention where the quadratic norm of $\vec{\alpha}$ is $1$.

\end{document}